\documentclass[reprint, prb, floatfix, aps, superscriptaddress]{revtex4-2}
\pdfoutput=1
\usepackage{amssymb, latexsym}
\usepackage{amsmath}
\usepackage{graphicx}
\usepackage{wrapfig}
\usepackage{chemformula}
\usepackage{mathtools}
\usepackage{tabularx}
\usepackage{color}
\usepackage{dcolumn}
\usepackage{bm}

\newcommand{\lafxy}{\lambda_{\text{AF}}^{xy}}
\DeclareMathOperator{\diag}{diag}
\renewcommand{\mathbf}[1]{\bm{#1}}

\begin{document}

\title{Tuning the confinement potential between spinons in the Ising chain 
\ch{CoNb2O6} using longitudinal fields and quantitative determination of the 
microscopic Hamiltonian}
\author{Leonie Woodland}
\affiliation{Clarendon Laboratory, University of Oxford Physics Department, 
Parks Road, Oxford OX1 3PU, UK} 
\author{David Macdougal}
\affiliation{Clarendon Laboratory, University of Oxford Physics Department, 
Parks Road, Oxford OX1 3PU, UK} 
\author{Ivelisse~M.~Cabrera} 
\affiliation{Clarendon Laboratory, University of Oxford Physics Department, 
Parks Road, Oxford OX1 3PU, UK} 
\author{Jordan~D.~Thompson} 
\affiliation{Clarendon Laboratory, University of Oxford Physics Department, 
Parks Road, Oxford OX1 3PU, UK}
\author{D. Prabhakaran} 
\affiliation{Clarendon Laboratory, University of Oxford Physics Department, 
Parks Road, Oxford OX1 3PU, UK} 
\author{Robert~I. Bewley} 
\affiliation{ISIS Facility, Rutherford Appleton Laboratory, Chilton, Didcot OX11 
0QX, UK}
\author{Radu Coldea}
\affiliation{Clarendon Laboratory, University of Oxford Physics Department, 
Parks Road, Oxford OX1 3PU, UK}

\date{\today}

\begin{abstract}
The Ising chain realizes the fundamental paradigm of spin fractionalization, 
where locally flipping a spin creates two domain walls (spinons) that can 
separate apart at no energy cost. In a quasi-one-dimensional system, the 
mean-field effects of the weak three-dimensional couplings confine the spinons 
into a Zeeman ladder of two-spinon bound states. Here, we experimentally tune 
the confinement potential between spinons in the quasi-one-dimensional Ising 
ferromagnet \ch{CoNb2O6} by means of an applied magnetic field with a large 
component along the Ising direction. Using high-resolution  single crystal 
inelastic neutron scattering, we directly observe how the spectrum evolves from 
the limit of very weak confinement at low field (with many closely-spaced bound 
states with energies scaling as the field strength to the power 2/3) to very 
strong confinement at high field (where it consists of a magnon and a dispersive 
two-magnon bound state, with a linear field dependence). 
At intermediate fields, we explore how the higher-order bound states disappear 
from the spectrum as they move to higher energies and overlap with the 
two-particle continuum. By performing a global fit to the observed spectrum in 
zero field and high field applied along two orthogonal directions, combined with 
a quantitative parameterization of the interchain couplings, we propose a 
refined single-chain and interchain Hamiltonian that quantitatively reproduces 
the dispersions of all observed modes and their field dependence. 
\end{abstract}

\maketitle

\section{Introduction}
\begin{figure}
\includegraphics[width=0.5\textwidth]{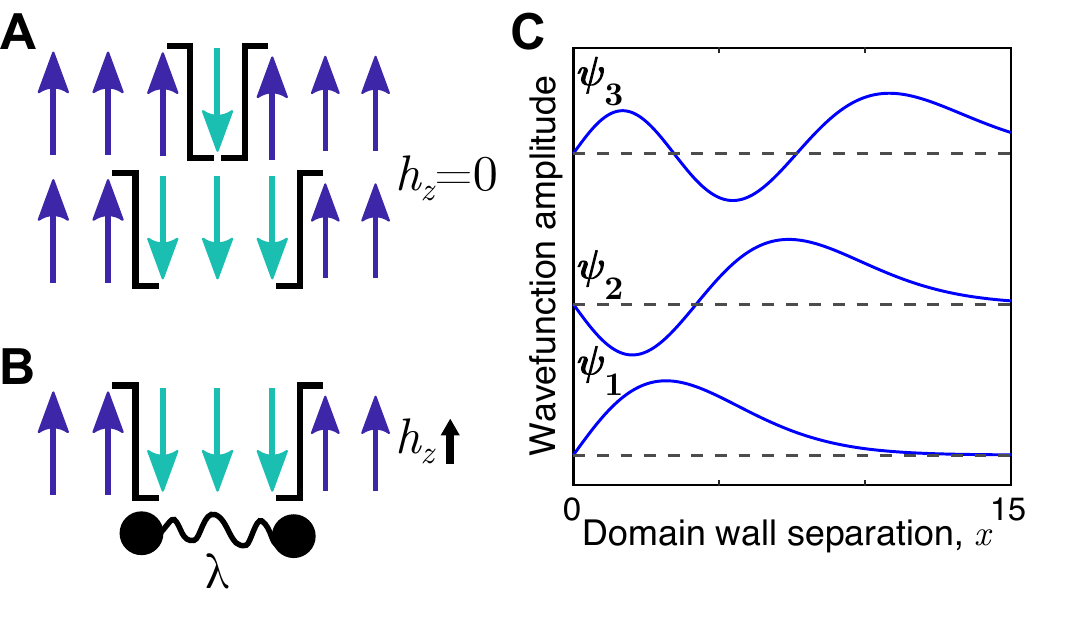}
\caption{A) Domain walls (solid lines) can separate apart at no energy cost in a 
ferromagnetic Ising chain. B) In a finite longitudinal field $h_z$, there is an 
energy cost linear in the separation, as if the two domain walls interacted via 
a string tension $\lambda$ (curly line between solid dots), which stabilizes 
confinement bound states. C) In the weak confinement regime ($h_z\ll J$), many 
bound states exist, which are coherent superpositions of states with the two 
domain walls separated by many sites. Solid lines are the Airy wavefunctions for 
the first three bound states as per (\ref{E:PsiAiry}) for parameters 
$\lambda\mu/\hbar^2=0.072$, relevant for the longitudinal mean field due to 
interchain interactions in \ch{CoNb_2O_6} \cite{Coldea2010}. The domain wall 
separation, $x$, is in units of the  spacing along the chain $c/2$. 
}\label{F:weakschematic}
\end{figure}

Fractionalization of coherently propagating spin-flips into two or more 
quasiparticles is a phenomenon of much interest in condensed matter physics. 
While in two and higher dimensions such phenomena require highly frustrated 
interactions, in one dimension, reduced mean field effects lead to 
fractionalization even in unfrustrated systems. A canonical example is the Ising 
chain in tilted field, with a deceptively simple Hamiltonian   
\begin{equation}\label{E:Htiltedising}
\mathcal{H}= \sum_j-J S^z_jS^z_{j+1} - h_xS^x_j -h_zS^z_j,
\end{equation}
with $J>0$ the Ising exchange, and $h_x$ and $h_z$ applied transverse and 
longitudinal fields, respectively. Consider first the case $h_z=0$: starting 
from a ferromagnetic alignment of all spins along the Ising $z$ axis and 
flipping a single spin creates two domain walls. These can separate at no energy 
cost and move apart independently of each other in the presence of the 
transverse field \cite{Lieb1961,Pfeuty1970}, as illustrated in 
Fig.~\ref{F:weakschematic}A. Therefore, a local spin flip, created for example 
in a neutron scattering process, fractionalizes into a pair of domain wall 
quasiparticles (spinons, also referred to elsewhere as solitons or kinks). 
However, in the presence of a finite {\em longitudinal} field $h_z>0$, domain 
walls are no longer free but have an attractive interaction, because there is an 
energy cost proportional to $h_z$ that increases linearly with their separation. 
The longitudinal field acts as an effective string tension between the domain 
walls, not allowing them to separate but confining them into bound states, 
realizing a one-dimensional analogue of the confinement of quarks into mesons 
\cite{McCoy1978}. By tuning the strength of the longitudinal field $h_z$, one 
could then explore the crossover from the regime of weak confinement (with $h_z$ 
a small perturbation on the scale of $J$) where many closely-spaced bound states 
are expected, with energy separation predicted to scale as a power-law 
$h_z^{2/3}$, to the regime of strong confinement ($h_z$ comparable to $J$), 
where, depending on the relative sizes of $J$, $h_z$ and $h_x$ and on what other 
subleading exchange terms may be present in the spin Hamiltonian beyond the 
minimal model in (\ref{E:Htiltedising}), only one or at most two bound states 
are expected, with their energies scaling linearly with $h_z$. The motivation 
behind the present studies was to explore the manifestation of this physics 
experimentally in a material where an external magnetic field can be used to 
tune the 
longitudinal field $h_z$
to cover the full range from weak to strong confinement.   

\begin{figure}
\centering
\includegraphics[width=0.3\textwidth]{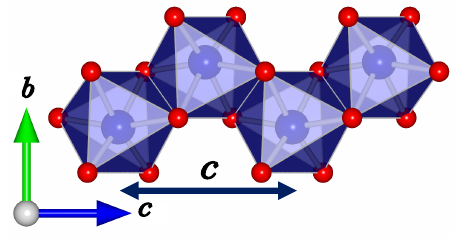}
\caption{A zigzag magnetic chain in \ch{CoNb_2O_6}, showing \ch{Co^{2+}} ions 
(blue) inside edge-sharing octahedra (blue shading) of \ch{O^{2-}} ions (red). 
}
\label{F:structure}
\end{figure}

The material \ch{CoNb_2O_6} is considered to be an excellent experimental 
realization of a ferromagnetic Ising chain with a low enough exchange energy 
scale that the full phase diagram in magnetic field is experimentally accessible 
\cite{Coldea2010,Kinross2014xe,Liang2015,Amelin2020ov,
Amelin2022,Morris2014ux,Wheeler2007}. 
It displays a quantum phase transition in transverse field, from an ordered 
phase to a quantum paramagnet, and around the critical point it displays the 
expected universal properties, such as evidence for an emergent E8 spectrum  
\cite{Coldea2010,Amelin2020ov}. The magnetic ions are \ch{Co^{2+}}, arranged 
into zigzag chains running along the crystallographic $c$-direction, with the 
zigzag in the $b$-direction, as shown in Fig.~\ref{F:structure}. The chains form 
a distorted triangular lattice in the $ab$-plane and the crystal structure is 
orthorhombic (space group $Pbcn$) with lattice parameters ${a=14.1337}$~\AA, 
$b=5.7019$~\AA ~ and $c=5.0382$~\AA ~ at 2.5~K \cite{Heid1995bt}. The 
combination of crystal field effects and spin-orbit coupling leads to an 
effective $S=1/2$ Kramers doublet ground state with strong Ising-like character, 
which is separated from the next lowest Kramers doublet by 30~meV 
\cite{Ringler2022}. Weak interactions between chains stabilize magnetic order at 
low temperatures, and below 1.97~K, the spins are ferromagnetically aligned 
along each zigzag chain due to a dominant nearest-neighbour Ising exchange, with 
an antiferromagnetic pattern between chains \cite{Heid1995bt,Mitsuda1994}. The 
magnetic moments lie in the $ac$-plane \cite{Maartense1977ef,Scharf1979gi}, at 
an angle $\gamma=30^{\circ}$ to the $c$-axis \cite{Heid1995bt}, which we take to 
be the local Ising $z$-axis; a field applied along the $b$-axis is then 
transverse to the Ising axes of all spins. The ferromagnetic nature of the 
dominant interactions makes \ch{CoNb_2O_6} an ideal candidate for experimental 
tuning of the confinement potential between spinons. 

Here, we report high resolution single crystal inelastic neutron scattering 
(INS) measurements that probe the full wavevector-dependence of the spectrum as 
a function of magnetic field applied along the crystallographic $a$-axis, with a 
large longitudinal (along Ising axis) component. Our 
results complement earlier terahertz (THz) spectroscopy measurements for the 
same applied field 
direction, which probed the spectrum at the zone centre 
(wavevector transfer $\mathbf{Q}=0$)
\cite{Morris2014ux,Kunimoto1999aa}, as well as 
THz spectroscopy and INS measurements 
previously taken on the Ising material {\ch{CoCl2 * 2 H2O}} where the interchain 
couplings are more significant \cite{Torrance1969b,Hansen2022aa}.
In particular, we parameterize the evolution of the bound state spectrum with 
field and find good agreement with scaling laws expected in the regimes of weak 
and strong confinement, corresponding to low and high applied fields, 
respectively. We also parameterize the effects of the interchain couplings on 
the high-field dispersive modes for fields along both $a$ and $b$ to extract 
pure 
one-dimensional (1D) dispersion relations. We then compare those with exact 
diagonalization (ED) calculations to arrive at a refined spin Hamiltonian that 
quantitatively reproduces all the features seen in the full wavevector 
dependence of the inelastic neutron scattering data, across a wide range of 
applied fields. 

The rest of this paper is organized as follows. Section \ref{S:experimental} 
outlines the inelastic neutron scattering experiments and Sec.~\ref{S:results} 
presents the experimental results for the spectrum in field applied along $a$ 
across a wide range of field strengths from weak confinement 
(\ref{S:weakconfinement}) to strong confinement (\ref{S:strongconfinement}), 
with the evolution of the high-order bound states presented in 
Sec.~\ref{S:crossover}. Next, in Sec.~\ref{S:hamiltonian}, we propose a 
Hamiltonian for the in-chain interactions and refine the 
parameters by comparison to the observed dispersions at zero field and high 
field along two orthogonal directions ($a$ and $b$). Sec.~\ref{S:conclusions} 
contains our conclusions. The Appendices contain the characterization and 
parameterization of the effects of the interchain interactions, as well as 
comparisons between the data 
and other parameter sets.

\section{Experimental details}\label{S:experimental}
Inelastic neutron scattering measurements of the magnetic excitation spectrum 
were performed using the direct geometry time-of-flight spectrometer LET at the 
ISIS facility \cite{LET}. The sample was a large single crystal (4.59~g) of 
\ch{CoNb_2O_6} 
grown by the floating-zone technique, as described in \cite{Prabhakaran2003}, 
and mounted in the $(0kl)$ horizontal scattering plane, where the wavevector 
transfer $\bm{Q}$ is labelled as $(hkl)$ in reciprocal lattice units of the 
structural orthorhombic unit cell, so 
${\bm{Q}}=2\pi\left(\frac{h}{a},\frac{k}{b},\frac{l}{c}\right)$. 
The sample was mechanically fixed in place using copper brackets so that it 
would not move or rotate due to the torques from the applied field. The sample 
was 
cooled using a dilution refrigerator insert and all data were collected below 
0.14~K. A magnetic field up to 9~T was applied vertically, along the 
crystallographic $a$-axis, which has a longitudinal field component 
$h_z=g_z\mu_BB\sin\gamma$, where $B$ is the externally applied field magnitude. 

LET was operated to measure simultaneously the inelastic scattering of incident 
neutrons with energies of $E_i=2.46$, 4.30 and 9.33~meV; the measured energy 
resolutions (full width half maximum, FWHM) on the elastic line were 0.038(1), 
0.085(1), and 0.274(2)~meV, respectively. To obtain the overview of the field 
dependence of the spectrum, the scattering was measured at each field for one 
fixed sample orientation chosen such that data projected along the chain 
direction covered a large part of the along-chain Brillouin zone. For these 
measurements, the sample was aligned with the (010) axis at an angle of 
$8^{\circ}$ to the incident beam direction, and scattering was measured for a 
typical counting time of 3 hours per field at an average beam current of 40 
$\mu$A of protons on target. In addition to the single 
orientation measurements, multi-angle (Horace) scans
were performed to obtain a full 
four-dimensional data set of scattering intensity as a function of three 
momentum directions and energy at two representative fields, 1.5 and 8~T. For 
these, the sample was rotated about the vertical $a$ axis in steps of 
$2^{\circ}$ over an angular range of $88$ and $108^{\circ}$, with 11 and 17 
minutes counting per step, respectively. 
Additional Horace scans were 
collected in a separate experiment 
on LET at 0.14 K in a field of 9~T~$\parallel b$ on the same crystal as in 
\cite{Cabrera2014yj} mounted in the $(h0l)$ horizontal scattering plane. For 
these measurements, the incident energies used were $E_i= 2.14$, 4.02 and 
10.17~meV, and the angular range was 145$^\circ$ covered in steps of 1$^\circ$ 
with 6 minutes counting per step. 
The Horace scans were used to 
quantitatively parameterize the interchain dispersions, as detailed in 
Appendix~\ref{A:highfieldinterchain}. The raw time-of-flight neutron data were 
converted to 
scattering intensities $S(\bm{Q},\omega)$ using \textsc{mantid} 
\cite{Arnold2014}, and the resulting data were then analysed using the 
\textsc{horace} \cite{Ewings2016} and \textsc{mslice} \cite{MSlice} packages.

\section{Results}\label{S:results}

\subsection{Weak confinement regime}\label{S:weakconfinement}

\begin{figure*}
\includegraphics[width=\textwidth]{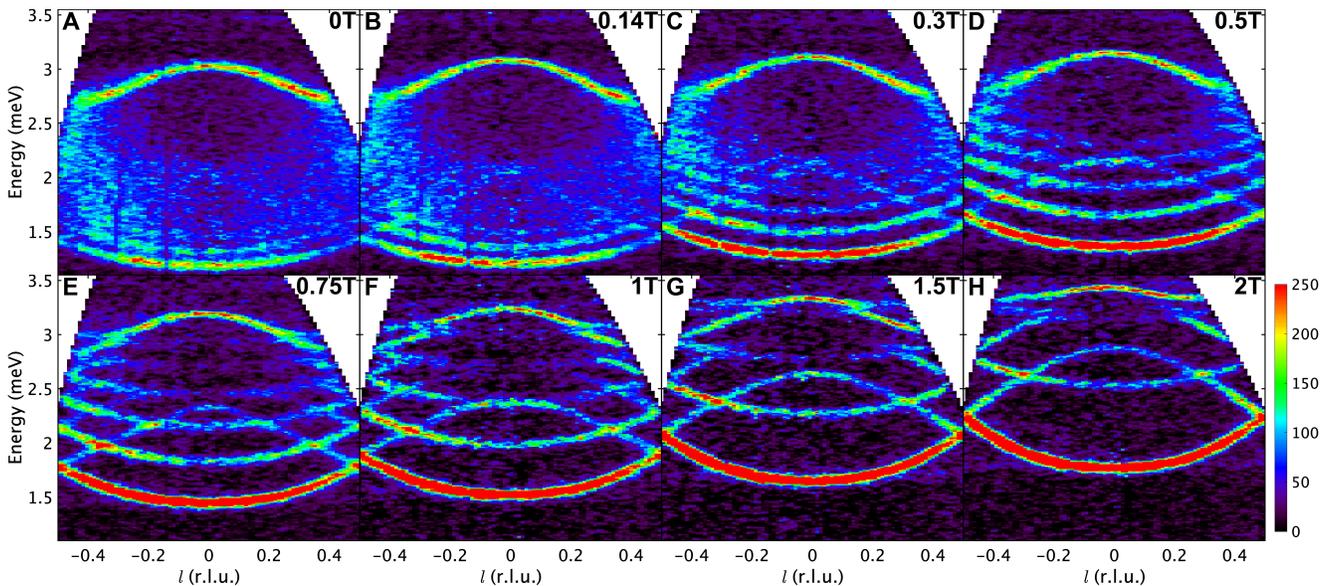}
\caption{INS spectrum as a function of field from 0 to 2~T $\parallel a$, in the 
weak to intermediate confinement regimes, with field increasing from left to 
right and top to bottom. Color is the raw neutron scattering intensity (i.e., no 
background has been subtracted) on an 
arbitrary scale, collected using a high-resolution configuration 
($E_i=4.30$~meV) for a fixed sample orientation, with wavevectors projected 
along the chain direction.}\label{F:weakconfinement}
\end{figure*}

In the limit of small longitudinal field $h_z\ll J$, the confinement of domain 
walls into bound states can be captured via a Schr\"{o}dinger equation for the 
domain wall separation distance $x$ in the continuum limit 
\cite{McCoy1978,Fogedby1978,Rutkevich2008aa}
\begin{equation}\label{E:Airy}
-\frac{\hbar^2}{\mu}\frac{d^2\psi_n(x)}{dx^2}+\lambda|x|\psi_n(x)=(m_n-2m_0)
\psi_n(x),
\end{equation}
where $\mu$ is the reduced mass for the domain wall pair, $\lambda=2h_z/c$ is 
an effective string tension between the domain walls, $c/2$ is the spacing of 
sites along the chain, and $m_0$ is the energy cost to create a single domain 
wall. 
For the Hamiltonian (\ref{E:Htiltedising}), $m_0=J/2-h_x$ \cite{Sachdev1999}. 
Eq.~(\ref{E:Airy}) can be recast as the Airy equation, giving energy eigenvalues 
\begin{equation}\label{E:Eairy}
m_n=2m_0+\left(\frac{\hbar^2}{\mu}\right)^{1/3}\lambda^{2/3}z_n,
\end{equation}
where $-z_n$ are the zeros of the Airy $\operatorname{Ai}$ function; the 
corresponding eigenstates 
are 
\begin{equation}\label{E:PsiAiry}
{\psi_n(x)=\operatorname{Ai}\left(\left(\frac{\mu}{\hbar^2}\lambda\right)^{1/3}
\left(x-\frac{m_n-2m_0}{\lambda}\right)\right)}.\
\end{equation}
The wavefunctions for the lowest three bound states are sketched in 
Fig.~\ref{F:weakschematic}C.
This shows that the bound states in the weak confinement regime 
are expected to be superpositions of many states, with finite  amplitude even 
for states with large separations between 
domain walls.

Eq.~(\ref{E:Airy}) is applicable for \ch{CoNb_2O_6} even in the absence of an 
externally 
applied field. This is because i) there is a finite longitudinal (internal) mean 
field 
due to the weak interchain interactions $h_z=2J\lambda_{\text{MF}}\langle 
S^z\rangle$, 
with $J$ and $\lambda_{\text{MF}}$ defined in Sec.~\ref{S:propham} and $\langle 
S^z\rangle$ 
the expectation value of the spin component along the Ising axis; and ii) there 
are terms 
in the spin exchange Hamiltonian beyond the dominant Ising exchange (as will be 
shown 
in Sec.~\ref{S:hamiltonian}) that lead to domain wall propagation, so a finite 
$\mu$.  
The bound state spectra observed in previous INS \cite{Coldea2010} and THz 
spectroscopy 
experiments \cite{Morris2014ux,Amelin2020ov}  have shown good agreement with the 
predictions of (\ref{E:Airy}) with $h_z$ interpreted as the interchain 
longitudinal mean field, which has the same magnitude for all sites in the 
zero-field antiferromagnetic phase. Zero-field data collected in the current 
experimental setting are shown in Fig.~\ref{F:weakconfinement}A, where the three 
lowest bound states
are clearly visible at the lowest energies. Also notable is the sharp mode at 
the top of the spectrum with a dispersion curving the opposite way to the lower 
modes. This mode, which is stabilized by a different mechanism from the 
low-energy confinement bound states, is a single-spin-flip bound state 
stabilized by a subleading term in the spin Hamiltonian of the form 
$-J\lambda_{S}\sum_j\left(S_j^+S_{j+1}^-+S_j^-S_{j+1}^+\right)/2$. This XY 
exchange term allows a single-spin-flip state (i.e., two domain walls on 
adjacent 
sites) to hop between nearest-neighbour sites along the chain as a single entity 
and this state was therefore dubbed a kinetic bound state \cite{Coldea2010}. 

The zigzag nature of the magnetic chains illustrated in Fig.~\ref{F:structure} 
leads to a doubling of the unit cell along the chain direction compared to 
straight chains, therefore a doubling of the number of magnetic modes, so each 
``primary'' mode with dispersion $\omega(h,k,l)$ has a ``shadow'' version with 
dispersion $\omega(h,k,l+1)$ obtained from the ``primary'' version by Brillouin 
zone folding. 
In zero field, in the approximation of decoupled chains, the intensities of 
these modes are proportional to $\cos^2(2\pi k \zeta)$ and $\sin^2(2\pi k 
\zeta)$
respectively, where the magnetic ions alternate in position along the $b$ 
direction as $\pm\zeta b$ between consecutive sites along the zigzag chain 
($\zeta=0.165$) \cite{Cabrera2014yj}. In Fig.~\ref{F:weakconfinement}A, the top 
mode (the kinetic bound state) is a shadow mode with finite intensity only 
because of finite $k$, whereas all low-energy confinement modes are primary. 
Throughout this paper, we refer to the $n$th lowest energy (primary) mode as 
$m_n$, in both the weak confinement and strong confinement regimes. Modes with 
the same label in these two regimes have very different character, but there is 
a smooth crossover between the two regimes.

\begin{figure*}
\includegraphics[width=\textwidth]{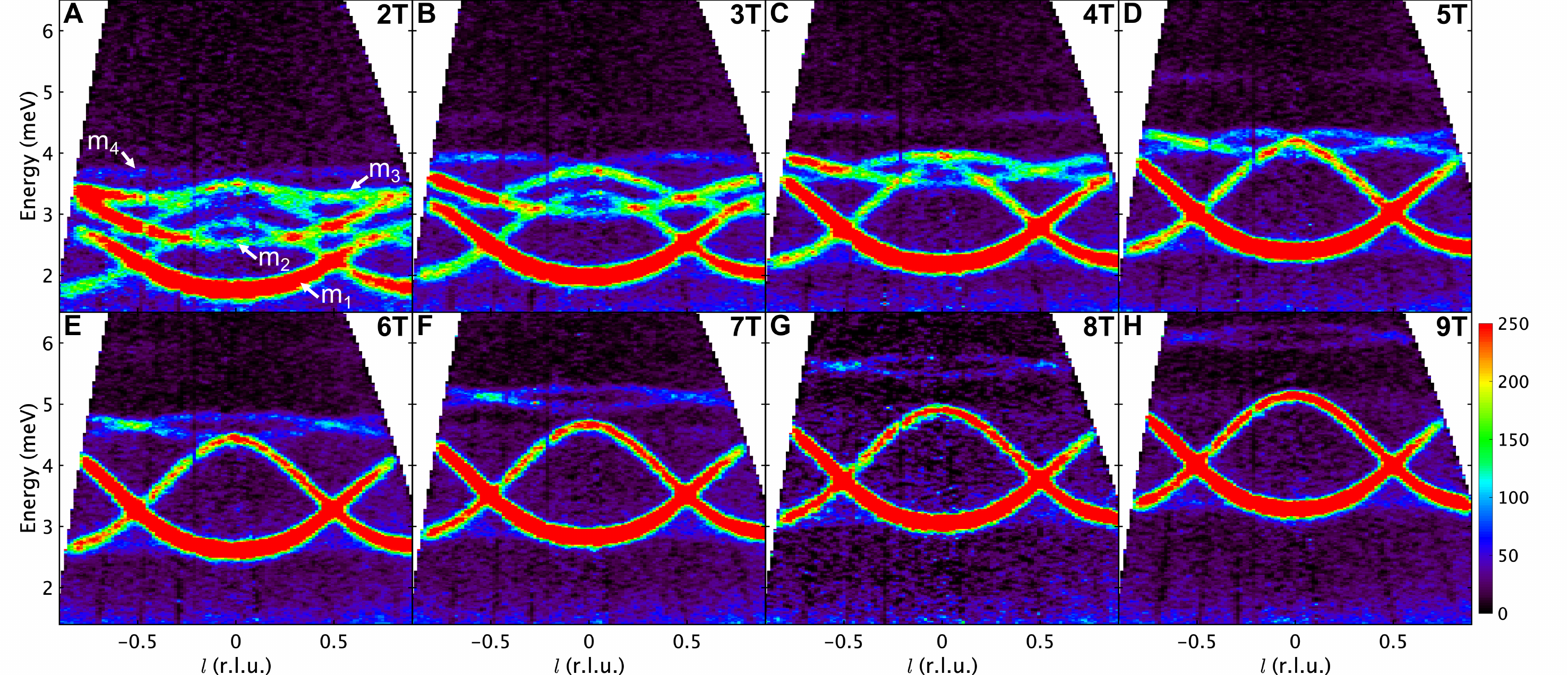}
\caption{INS spectrum as a function of field from 2 to 9~T $\parallel a$, 
covering the intermediate to strong confinement regimes, with field increasing 
from left to right and top to bottom. Color is the raw neutron scattering 
intensity on an arbitrary scale (different from Fig.~\ref{F:weakconfinement}), 
collected using a high-coverage coarser-resolution configuration 
($E_i=9.33$~meV) for the same fixed sample orientation as in 
Fig.~\ref{F:weakconfinement}, with wavevectors projected along the chain 
direction. Panel A is at the same field as the higher-resolution data in 
Fig.~\ref{F:weakconfinement}H.}\label{F:strongconfinement}
\end{figure*}

The remaining panels in Fig.~\ref{F:weakconfinement} show the evolution of the 
spectrum upon increasing field. It is known that low fields applied along $a$ 
induce a series of spin-flip transitions such that above a threshold field of 
0.14~T, all spin components along $a$ are parallel to the applied field 
\cite{Heid1997574,Weitzel2000ei,Kobayashi2016}. All finite field INS data were 
therefore collected at fields at and above this threshold field to ensure all 
spin sites experience the same magnitude longitudinal and transverse fields. 
Upon increasing applied field, all bound states move up in energy, with the 
higher order ones moving at a faster rate. This is because the higher order 
bound states contain in their superposition more weight for states with domain 
walls further apart [see Fig. \ref{F:weakschematic}C], so with more spins 
flipped opposite to the applied field direction, and therefore higher Zeeman 
energy or higher effective $g$-factor. Since the higher modes increase in energy 
more quickly, the relative energy separation between adjacent modes increases 
and modes become better resolved, such that whilst at 0.14~T (panel B) only 
$m_1$ to $m_3$ are resolved, at 0.5~T (panel D), $m_1$ to $m_7$ can be resolved. 
The kinetic bound state, however, increases in energy relatively slowly since it 
is a single-spin-flip state, such that the confinement bound states move past it 
as field is increased. Upon increasing field, the extent of the spectrum covers 
a progressively wider energy range and so, to probe the full spectrum above 2~T, 
we use a higher incident energy, coarser-resolution configuration with data 
plotted in Fig.~\ref{F:strongconfinement}. Data at 2~T are shown in both 
configurations, [compare Fig.~\ref{F:weakconfinement}H and 
Fig.~\ref{F:strongconfinement}A]. At 2~T (panel A), modes $m_1$ to $m_4$ (faint 
horizontal line near 3.7 meV) are clearly resolved, but, upon further increasing 
field, modes $m_3$ and $m_4$ become progressively fainter and above 6~T, in the 
strong confinement regime, only $m_1$ and $m_2$ remain (panels F to H). In the 
following we first discuss the strong confinement regime, then discuss in detail 
the intermediate field regime.

\subsection{Strong confinement regime}\label{S:strongconfinement}

\begin{figure}
\includegraphics[width=0.44\textwidth]{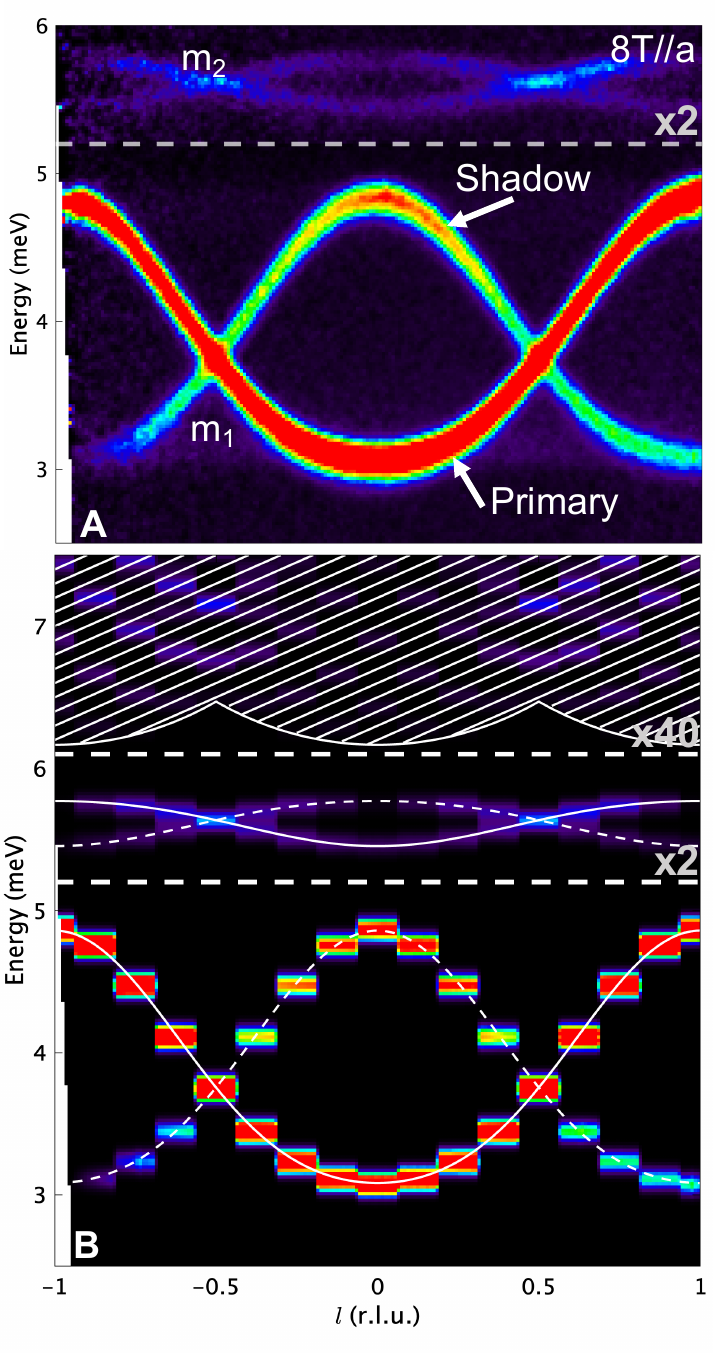}
\caption{A) INS spectrum at 8~T, deep in the strong confinement regime, obtained 
by integrating a multi-angle scan over the 
transverse wavevector range $|h| 
\leq 1.5$ and $|k| \leq 1.5$. Intensities above the dotted line have been scaled 
up by a factor of 2 to make them more clearly visible. The incident energy used 
is the same as in Fig.~\ref{F:strongconfinement}. B) Calculated intensity 
$S(\mathbf{Q},\omega)$ as defined in (\ref{E:Sqw}) for the single-chain 
Hamiltonian in (\ref{E:propHam}), where interchain interactions are included in 
a mean-field approximation as per (\ref{E:Hmf_a}), calculated using ED on a 
chain of $n=16$ ($n/2$ zigzag unit cells) sites with periodic boundary 
conditions. ED gives the spectrum for a discrete set 
of wavevectors spaced by $\Delta l=2/n$; for visualization purposes, the 
spectrum is plotted as constant over the interval $\Delta l$ around each 
discrete wavevector. The ED results have been convolved with a Gaussian 
in energy of FWHM 0.16~meV, representative of the experimental resolution 
in this energy range and obtained by a fit to the observed peak lineshapes. The 
white curves are the best fit 
single-chain dispersion relations with solid and dashed curves showing primary 
and shadow 
modes (obtained via $l \rightarrow l+1$ translation) respectively. The hatched 
region represents the $2m_1$ continuum obtained by assuming that quasiparticles 
do not interact. Note that $m_2$ is below the $2m_1$ continuum at all 
wavevectors. The $m_2$ and $2m_1$ intensities have been multiplied by factors of 
2 and 40, respectively.  
}\label{F:8Tdata}
\end{figure}

The INS spectrum at 8~T, representative of the strong confinement regime, is 
shown in Fig.~\ref{F:8Tdata}A. Two dispersive modes, lower energy $m_1$ and 
higher energy $m_2$, are seen, each with both a (stronger intensity) primary and 
a (weaker intensity) shadow version (obtained from the primary via an $l 
\rightarrow l+1$ translation). In this regime, $m_1$ is a magnon mode, a 
coherently propagating single spin flip, as expected for the lowest energy 
excitation in a field-polarized paramagnet. Meanwhile, the $m_2$ mode can be 
understood as a dispersive two-spin-flip state, i.e. a bound state of two $m_1$ 
magnons. 
To understand why a two-magnon state can still exist in this limit of high 
fields, and why it is dispersive, albeit with a much suppressed bandwidth 
compared to  $m_1$, it is insightful to consider a minimal Ising-like XXZ model 
in longitudinal field
\begin{align}
{{\cal{H}}_{\rm XXZ}}=-J&\sum_j \left[ S^z_jS^z_{j+1} + \lambda_S(S^x_jS^x_{j+1} 
+S^y_jS^y_{j+1})\right] \nonumber\\
-&\sum_jh_zS^z_j.
\label{E:XXZ}
\end{align}
The spectrum can be solved exactly \cite{Orbach1958} and contains, in addition 
to conventional one-magnon ($m_1$) excitations, also a two-magnon ($m_2$) bound 
state, with dispersion relations
\begin{align}
m_1(l) & =  h_z+J(1-\lambda_S\cos \pi l) \nonumber\\
m_2(l) & = 2h_z+ J\left[1-\frac{\lambda_S^2}{2}(\cos \pi l + 1)\right] 
\label{E:XXZ_w1w2}
\end{align}
and (un-normalized) wavefunctions, expressed in wavevector units relevant here, 
as
\begin{equation}\label{E:m1}
    |m_1\rangle=\sum_j e^{i\pi 
lj}|\dots\uparrow\uparrow\uparrow\downarrow_j\uparrow\uparrow\uparrow\dots
\rangle
\end{equation}
and
\begin{align}
    |m_2\rangle&=\sum_j e^{i\pi l 
j}\Bigl(|\dots\uparrow\uparrow\downarrow_j\downarrow\uparrow\uparrow\uparrow
\dots\rangle 
\nonumber\\
    &+ \lambda_S e^{i\pi l/2} \cos{\frac{\pi 
l}{2}}|\dots\uparrow\uparrow\downarrow_j\uparrow\downarrow\uparrow\uparrow\dots
\rangle 
+ \mathcal{O}(\lambda_S^2)\Bigr).\label{E:m2}
\end{align} 
For $m_1$, the dispersion comes at first order in the XY exchange $J\lambda_S$. 
For $m_2$, the hopping process occurs in two stages via an intermediate state in 
which the spin flips are separated by one site [second term in (\ref{E:m2})], 
therefore the hopping process is second order in the XY exchange, which leads to 
the dispersive term proportional to $\lambda_S^2$ in (\ref{E:XXZ_w1w2}). We 
propose that the $m_2$ state seen experimentally has the same qualitative 
content as the two-spin-flip bound state for the minimal model in (\ref{E:XXZ}), 
but with modifications appropriate for the relevant Hamiltonian, such that the 
$m_2$ dispersion relation depends on all $\lambda$'s in (\ref{E:propHam}), to be 
discussed later. The dynamical correlations calculated for this model are shown 
in Fig.~\ref{F:8Tdata}B and capture all key features of the dispersions and 
intensities of both $m_1$ and $m_2$ modes. 

Note that a two-spin-flip state cannot usually be seen in inelastic neutron 
scattering [it would not be observable for the Hamiltonian in (\ref{E:XXZ})], 
since the scattering intensity of a given state is determined by its overlap 
with a single spin flip created in a neutron scattering process. However, in the 
present case, the external field is applied along the crystallographic $a$-axis, 
which makes a finite angle $90^{\circ}-\gamma$ with the Ising direction, and so 
the field is not perfectly longitudinal but also has a finite transverse 
component. In addition, as we discuss in Sec.~\ref{S:hamiltonian}, there are 
also sub-leading exchange terms in the spin Hamiltonian that break spin 
conservation. Both the finite (large) transverse field and the (relatively much 
smaller) exchange terms in the Hamiltonian that break spin conservation lead to 
some mixing between states with different numbers of spin flips, allowing the 
predominantly two-spin-flip $m_2$ state to have a finite admixture of a 
single-spin-flip state and therefore to be visible in the present INS 
experiments. 

It is notable that the $m_2$ mode is expected to survive up to indefinitely 
large field (although the intensity would become progressively weaker as the 
mixing with the single-spin-flip state would progressively decrease upon 
increasing Zeeman energy). 
This is because the binding energy of $m_2$ relative to the $2m_1$ continuum for 
the minimal model in (\ref{E:XXZ}, \ref{E:XXZ_w1w2}) is $m_2-2m_1=-J$ in the 
limit $\lambda_S\rightarrow 0$,
i.e. the two magnons gain Ising energy if they are next to one another.
The fact that the $m_2$ mode is lower in energy than the $2m_1$ continuum means 
that it is not possible for it to decay, hence it survives as a sharp mode.

It is also important to note that at these high fields the $m_1$ state has 
non-negligible dispersion bandwidth perpendicular to the chain direction because 
of the 
finite interchain couplings; this is discussed further in 
Appendix~\ref{A:highfieldinterchain}. This has the consequence that the 
lineshape of the 
$m_1$ mode in Fig.~\ref{F:8Tdata}A appears artificially broadened because the 
data are integrated over a large range in the wavevector direction transverse to 
the chains. The interchain dispersion at high field is non-negligible because 
the hopping strength between chains for a single spin flip is first order in the 
interchain exchange. In contrast, at lower fields, even the lowest energy mode 
($m_1$) has multiple spin flips, as illustrated in Fig.~\ref{F:weakschematic}C 
for $\psi_1$, so the interchain dispersion is higher order and is thus much 
suppressed, such that it is essentially undetectable at low field.

We also note that the bottom of the $m_1$ dispersion is visibly flattened 
compared to a perfect sinusoid [see Fig.~\ref{F:8Tdata}A], indicating a double 
Fourier component along $l$, suggesting hopping to next-nearest-neighbour sites 
along the chain, which will be further discussed in Sec.~\ref{S:propham}. 

\subsection{Fate of higher-order bound states}\label{S:crossover}

\begin{figure*}
\includegraphics[width=0.85\textwidth]{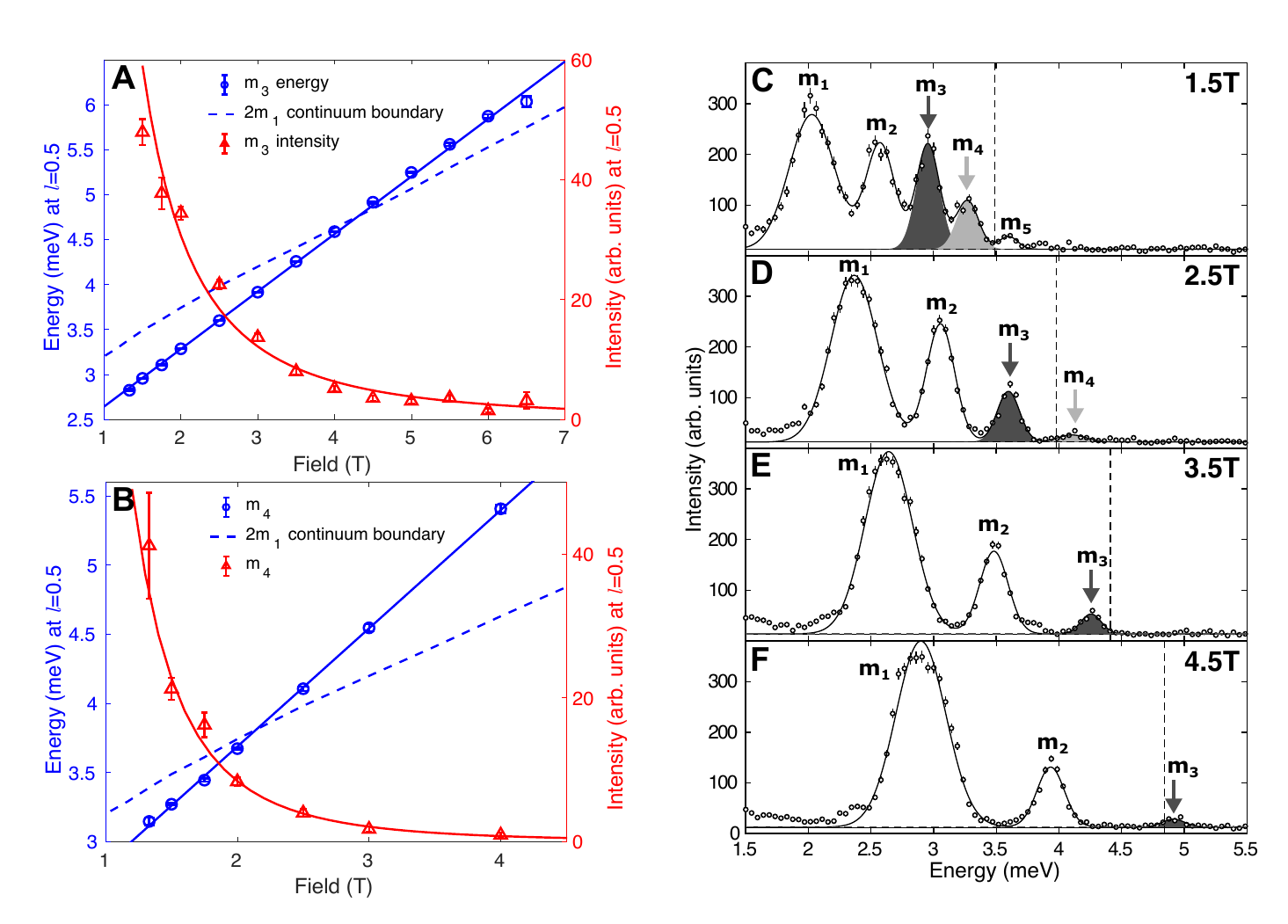}
\caption{A,B) Energies (blue symbols, left axes) and integrated intensities (red 
symbols, right axes) deduced from multi-Gaussian fits of the $m_{3}$ (A) and 
$m_4$ (B) mode at $l=0.5$, extracted from cuts from data in the same 
configuration as in Fig.~\ref{F:strongconfinement}. Solid lines are guides to 
the eye and the dashed lines show the estimated $2m_1$ continuum lower boundary 
at the same $l$-value. C-F)
Energy scans showing the field dependence of the various modes (integrated over 
$0.4 < l < 0.6$), solid lines are fits to multi-Gaussian peaks. Intensities are 
raw neutron scattering intensities on an arbitrary scale (the same for all 
panels). Note that a clear signal at the expected $m_{3,4,5}$ energy is present 
even after crossing the $2m_1$ continuum lower boundary (dashed vertical line in 
panels C-F),  which has been calculated using fitted dispersion relations for 
the $m_1$ 
mode and assuming that $m_1$ particles do not interact with each 
other.}\label{F:m3m4intensities}
\end{figure*}

\begin{figure}
\includegraphics[width=0.469\textwidth]{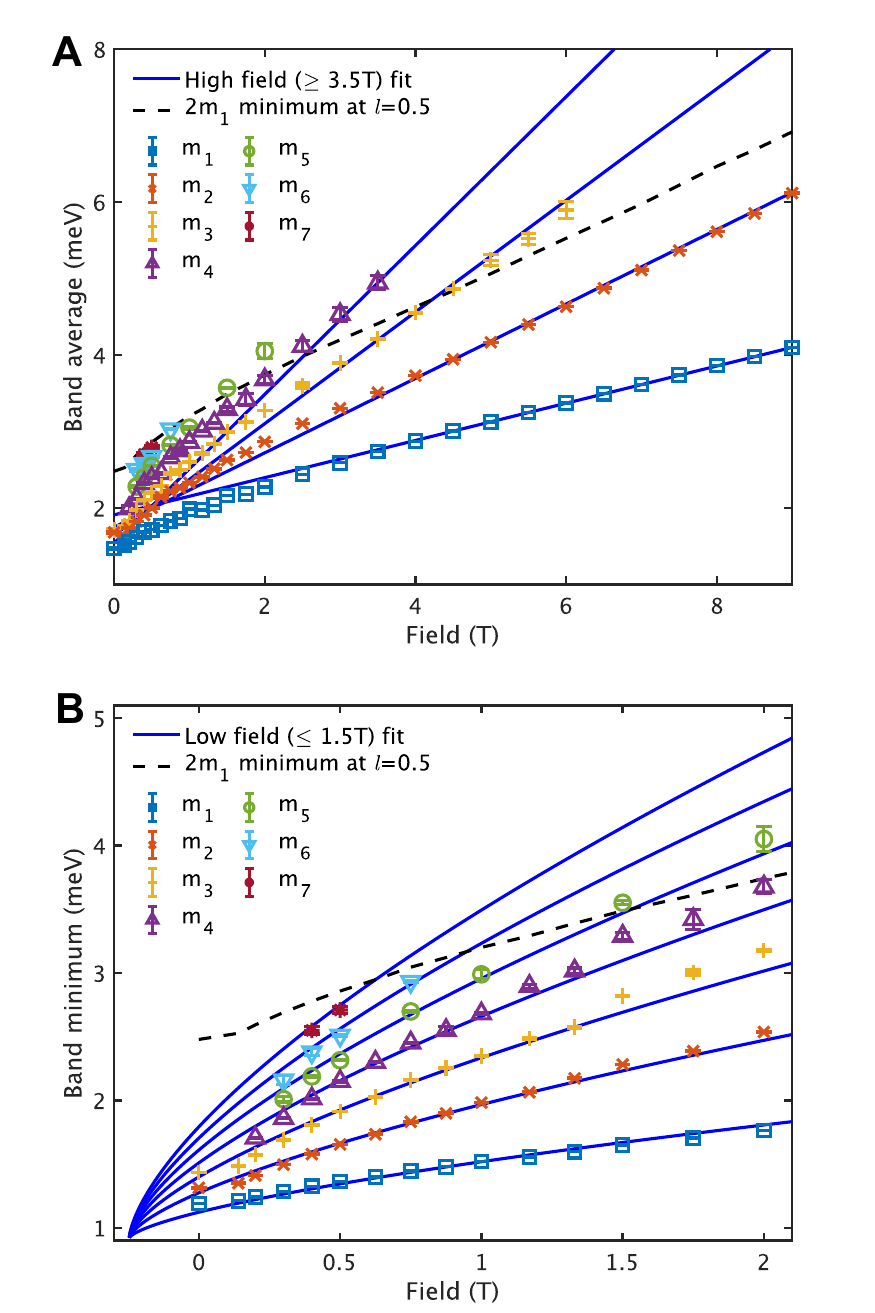}
\caption{A) Band averages fit (solid lines) to a form $E_n=ng\mu_BB+E_{0n}$ 
where $n$ labels modes, $g=4.20(2)$ is an effective $g$-factor (for fields along 
$a$) and $E_{0n}$ is an effective zero-field energy. The fit only includes data 
for $B\geq3.5$~T. B) One-dimensional band minima, corrected for the effects of 
interchain hopping, as a function of field in the low field regime. The band 
minima have been simultaneously fit (solid lines) to a form $E_n=\alpha 
z_n|B-B_0|^{2/3}+E_0$ as per (\ref{E:Airy}), where $-z_n$ are the zeroes of the 
Airy Ai function, $B_0=-0.24(2)$~T is a field offset due to the mean field 
effects of the antiferromagnetic interchain interactions, 
$\alpha=0.221(2)$~meV~T$^{-2/3}$ is a constant of proportionality related to 
$(\hbar^2/\mu)^{1/3}$, and $E_0=0.92(2)$~meV is the energy required to create a 
pair of domain walls. The fit only includes data in the field polarized 
paramagnetic regime for $0.14\leq B\leq1.5$~T; the zero field data were omitted 
from the fit as the magnetic structure is different from that above 0.14~T such 
that a different value of $B_0$ would be needed. In both panels, the dashed line 
shows the energy of the lower boundary of the $2m_1$ continuum at $l=0.5$ under 
the assumption that the quasiparticles do not 
interact.}\label{F:highlowfieldfit}
\end{figure}

The  evolution of the INS spectrum upon increasing field from 2 to 9~T is shown 
in Fig.~\ref{F:strongconfinement}. 
A notable feature is that, upon increasing field, the higher order bound states 
progressively become less dispersive and rapidly lose spectral weight, with the 
highest ones doing so at the fastest rate. In detail, at 2~T (panel A), $m_1$ 
and $m_2$ have similar intensities and bandwidths, $m_3$ is fairly flat and 
$m_4$ is very flat and fairly weak. Upon increasing field, $m_4$ moves up in 
energy and becomes weaker until it disappears between 4 and 4.5~T. At those 
fields, $m_3$ is completely flat and weaker in intensity, then becomes 
undetectable above 6.5~T.  

The field dependence of the $m_3$ mode is summarized in 
Fig.~\ref{F:m3m4intensities}A. The intensity (red triangles and right vertical 
axis) drops off very rapidly upon increasing field, decreasing by more than an 
order of magnitude between 1 and 4~T. In this field range, the $m_3$ energy 
(blue round symbols, left axis) is below the estimated $2m_1$ continuum lower 
boundary (dashed line), so the $m_3$ mode should be stable, with no mechanism 
for decay; nevertheless, its intensity drops off very fast upon increasing 
field. A qualitatively similar behaviour is observed for the $m_4$ mode (panel 
B): the intensity drop-off is an even steeper function of increasing field and 
again the intensity decreases very quickly even before 2~T, above which $m_4$ 
enters the $2m_1$ continuum. Within the sensitivity of the experiments, no 
systematic intensity change occurs for either of the $m_{3,4}$ modes as they 
enter the continuum and their signal could be followed even up to significantly 
higher fields, when the modes should be deep into the continuum energy range 
[see Fig.~\ref{F:m3m4intensities}C-F]. This might suggest that the matrix 
elements for the decay processes $m_{3,4}\rightarrow m_1+m_1$ are relatively 
weak, and that the dominant effect leading to the fast intensity drop-off with 
increasing field is the progressively reduced mixing of the single-spin-flip 
state into the wavefunction of the $m_{3,4}$ modes, as upon increasing field the 
number of spin flips becomes a progressively better approximation for a good 
quantum number. 
This picture is consistent with the results of ED calculations, which indicate 
that while there is some decay of the bound states when they overlap with the 
continuum, the predicted broadening is of order 0.01 meV, which is not 
resolvable within the experimental results presented here.

As well as studying the disappearance of the higher bound states, the crossover 
between the weak and the strong confinement regimes was quantitatively tested by 
doing fits to the expected field dependence of the mode energies. In the high 
field regime, a linear dependence of mode energy on field is expected. This is 
because, in this regime, the number of spin flips in the mode is approximately a 
good quantum number, with the $m_n$ mode containing states with $n$ spins 
flipped compared to the field-polarized state. Thus, the Zeeman energy is 
expected to scale linearly with both mode number and field. Indeed, the band 
averages are well fit by a linear dependence of energy on field and mode number, 
i.e., a fit where the gradients of the $m_2$, $m_3$ and $m_4$ lines are 
constrained to be 2, 3 and 4 times respectively the gradient of the $m_1$ line,  
as illustrated in Fig.~\ref{F:highlowfieldfit}A. A cutoff field of 3.5~T, 
assumed to be close enough to the high field limit, was used, and only data at 
fields above this were included in the fit, which was performed simultaneously 
to the four lowest energy modes.  The quality of this fit, especially for the 
$m_1$ and $m_2$ modes, also adds evidence that these are single-spin-flip and 
two-spin-flip modes respectively, i.e. derived from modes with wavefunctions 
given in (\ref{E:m1}) and (\ref{E:m2}). The fit in this regime used the band 
averages, as we have found that the band minimum in this regime is affected by 
the tilting of the local magnetization towards the field, which leads to a 
field-dependent bandwidth. This effect is quantitatively understood and is 
discussed in more detail in Sec.~\ref{S:otherfields}.

Fig.~\ref{F:highlowfieldfit}A also illustrates that the linear fit breaks down 
at low fields. This is expected, since at low field the number of spin flips 
ceases to be even approximately a good quantum number, and instead the weak 
confinement physics described in Sec.~\ref{S:weakconfinement} holds. In this low 
field regime, a fit was performed simultaneously to the band minima of all the 
modes, as shown in Fig.~\ref{F:highlowfieldfit}B, with the field dependence 
having a $2/3$ power law form, as per (\ref{E:Eairy}), and the spacing between 
the energy levels being constrained by the zeros of the Airy function. The band 
minima were corrected for the effects of interchain dispersion by fitting the 
experimentally-extracted dispersion points to a parameterized three-dimensional 
(3D) dispersion relation of the form given in Appendix~\ref{S:interchainlong}, 
then setting 
the interchain hopping terms in those parameterizations to zero to correct for 
the small energy shifts due to interchain dispersion. 
The good agreement obtained between the field dependence of the bound state 
energies and the different expected power-law behaviours in the two field limits 
in Fig.~\ref{F:highlowfieldfit} (solid lines) lends support to the proposal 
that, in \ch{CoNb2O6}, the interchain couplings are sufficiently small relative 
to the dominant in-chain exchange, and the Ising character sufficiently strong, 
that both the weak and the strong spinon confinement regimes are realized 
experimentally via tuning of the applied field. The crossover region is 
approximately between 1.5 and 3.5~T. We note that, according to the Hamiltonian 
and parameters refined in Sec.~\ref{S:hamiltonian}, this crossover region is 
where the kinetic term for domain walls and the Zeeman term are comparable, such 
that the continuum approximation used in (\ref{E:Airy}) for the weak confinement 
regime is no longer applicable. 

\section{Quantitative determination of the Hamiltonian}\label{S:hamiltonian}

The Horace scans collected at high field have enabled us to further 
refine the microscopic model for this system beyond the minimal in-chain 
Hamiltonian proposed in \cite{Fava2020}. In this section, we will describe the 
refined in-chain terms, as well as the fitting 
procedure, with details of the interchain interactions referred to 
Appendix~\ref{A:highfieldinterchain}. There, we revise and extend a minimal 
model of the interchain interactions previously proposed to explain the 
dispersions in large transverse field \cite{Cabrera2014yj}, such that we can 
consistently reproduce the high field dispersions for field along both $a$ and 
$b$ directions. 
 
The quantitative parameterization of the interchain dispersions
obtained in Appendix~\ref{A:highfieldinterchain} allows us to 
obtain effective 1D dispersion relations to which a single-chain Hamiltonian can 
be fit, 
which is done in Secs.~\ref{S:propham} and \ref{S:singlechainfitting}: 
Sec.~\ref{S:propham} introduces the proposed single-chain Hamiltonian and 
Sec.~\ref{S:singlechainfitting} describes the fitting method. In 
Sec.~\ref{S:otherfields}, we then demonstrate that this proposed Hamiltonian 
also quantitatively captures the behaviour of the magnetic excitations in 
\ch{CoNb2O6} away from the fields at which the fit was performed. We also show 
that the proposed Hamiltonian can account for the spectrum observed in THz 
spectroscopy in \cite{Morris2021}.

\subsection{Proposed single-chain Hamiltonian}\label{S:propham}

\begin{figure}
    \includegraphics[width=0.45\textwidth]{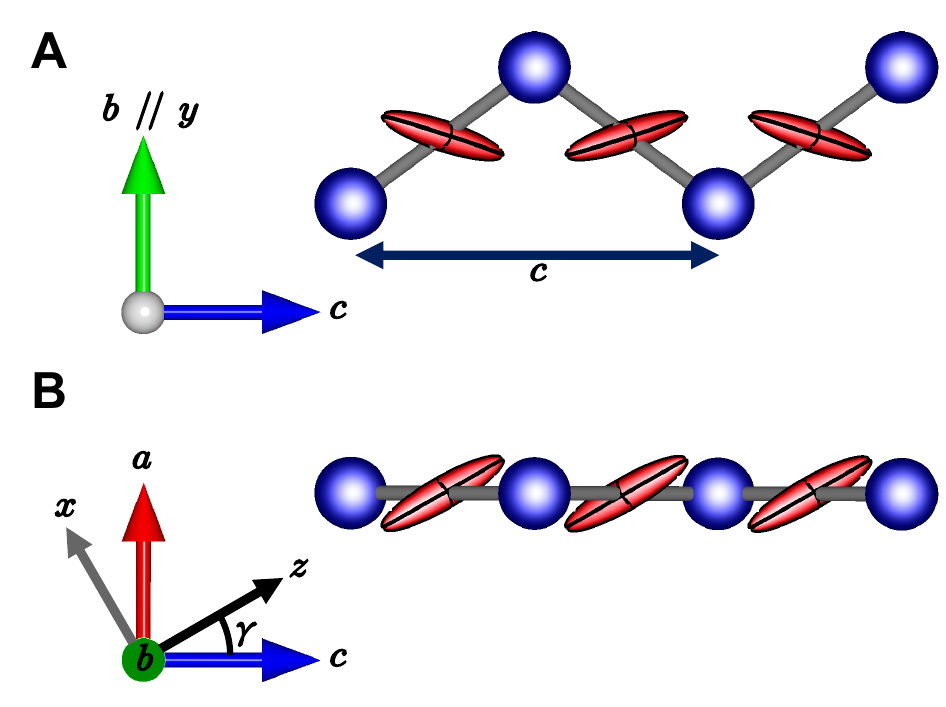}
    \caption{
    Graphical representation of the nearest neighbour part of the exchange 
Hamiltonian (\ref{E:propHam}) projected onto the $bc$ (A) and $ac$ (B) planes 
for a single chain. The blue spheres represent cobalt atoms while the red 
ellipsoids at the mid-point of each nearest-neighbour bond represent the 
exchange matrix on that bond. The principal axes of the ellipsoids correspond to 
the principal axes of the exchange matrix, with the lengths of the principal 
axes proportional to the absolute values of the relevant eigenvalues of the 
exchange matrix. A) Consecutive bonds along the chain are symmetry-related by 
the $c$-glide of the crystal structure (mirror in the $ac$ plane passing through 
the middle of each zigzag bond followed by translation by $\bm{c}/2$), which 
leads to the staggered orientations of the ellipsoids between consecutive bonds 
along the chain; this corresponds to the staggered exchange term $\lambda_{yz}$ 
in (\ref{E:Hmin}). B) The Ising $z$-axis is at an angle $\gamma$ to the $c$-axis 
in the $ac$ plane. The absolute signs of $\lambda_{yz}$ and angle $\gamma$ 
cannot be determined from neutron scattering measurements, only their magnitude, 
so only one of the four possible orientations of the exchange ellipsoids 
compatible with the experiments is shown. The other three options are obtained 
by mirroring the exchange ellipsoid for a reference bond in a plane passing 
through its centre, parallel to the $ab$, $bc$, or $ac$ crystallographic planes. 
The resulting orientation is then propagated via crystal symmetry operations 
onto all the other bonds, i.e. via the $c$-glide to obtain the other bonds on 
the same chain and via the $b$ (or $n$ glide) to relate bonds on the chain 
passing through the origin to bonds on the chain passing through the body-centre 
of the orthorhombic structural cell, as described in Fig.~\ref{F:Isingaxes}. }
 \label{F:exchange}
\end{figure}

The proposed single-chain Hamiltonian is an extension of one recently proposed 
on symmetry grounds \cite{Fava2020}, and it is convenient to write it as
\begin{equation}
\mathcal{H}_{\rm{single\: chain}} 
=\mathcal{H}_1+\mathcal{H}_2+\mathcal{H}_{\text{MF}}\label{E:propHam}
\end{equation}
where
\begin{align}
\mathcal{H}_1=J \sum_j [&-S_j^zS_{j+1}^z 
-\lambda_S\left(S_j^xS_{j+1}^x+S_j^yS_{j+1}^y\right)\nonumber\\
&+(-1)^j\lambda_{yz}\left(S_j^yS_{j+1}^z+S_j^zS_{j+1}^y\right)]\nonumber\\
-\mu_B\sum_j [&g_{x}B_xS_j^x + g_yB_yS_j^y + g_{z}B_zS_j^z]\label{E:Hmin}\\
\mathcal{H}_2=J\sum_j [&\lambda_{\text{AF}}S_j^zS_{j+2}^z 
+\lafxy\left(S_j^xS_{j+2}^x+S_j^yS_{j+2}^y\right) \nonumber\\
 &-\lambda_A\left(S_j^xS_{j+1}^x - S_j^yS_{j+1}^y\right)]\label{E:H2}
\end{align}
where $j$ runs over all sites on a single chain. Here, $xyz$ form an orthogonal 
right-handed coordinate system with $y$ along the crystallographic $b$-axis and 
$z$ 
defined to be the equilibrium spin direction in zero applied field, 
which for the Hamiltonian (\ref{E:propHam}) coincides with the direction along 
which the spin components have the largest exchange, i.e. the Ising axis. 

We will show that $\mathcal{H}_1$ is the \emph{minimal} Hamiltonian required to 
\emph{qualitatively} reproduce all key features of the the INS spectra seen in 
zero field, high field along $a$, and high purely-transverse field, while 
$\mathcal{H}_2$ is needed to \emph{quantitatively} capture all details of the 
spectra. 

\begin{figure*}
\includegraphics[width=\textwidth]{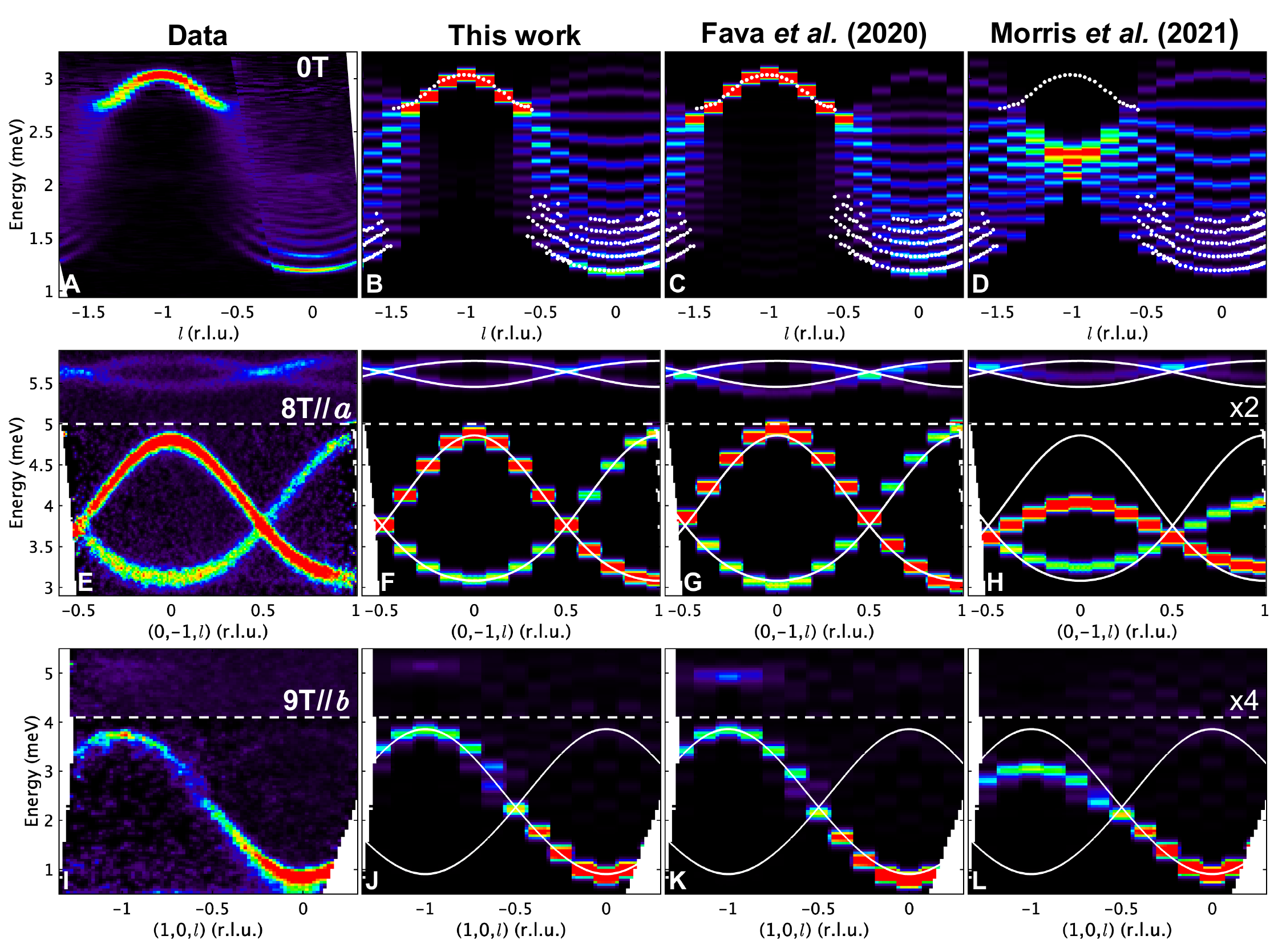}
\caption{
INS data in (A) zero field, (E) 8~T~$\parallel a$ and (I) 9~T~$\parallel b$ 
compared to dynamical correlations computed via ED for different models for a 
chain of 16 sites. Panel A is adapted from \cite{Coldea2010}. White dots in B-D 
show experimentally-extracted dispersion points from the data in A, solid white 
lines in F-H and J-L show the analytic parameterization of the experimental 
dispersion relations with interchain terms set to 0. The models used were:
the model 1D Hamiltonian in (\ref{E:propHam}) with parameters in 
Table~\ref{T:hamparams} (B, F, J);
the model refined in \cite{Fava2020} with $g_x$ and $g_z$ taken from 
Table~\ref{T:hamparams} (C, G, K); the model used in \cite{Morris2021} with 
$g_x$ and $g_z$ taken from Table~\ref{T:hamparams} (D, H, L).
In E-H, data below 5~meV (dashed line) are integrated over $|h|<1.5$ and 
$-1.1<k<-0.9$. Note that the shadow mode is the more intense mode here, as the 
range of transverse integration has been chosen in order to show the whole 
dispersion along $l$ while minimizing artificial broadening due to interchain 
dispersion. Data above 5~meV are integrated over $|h|<1.5$ and $|k|<1.5$ and 
intensities in this region have been multiplied by 2 to make the $m_2$ mode more 
clearly visible.
In I-L, data below 4.1~meV (dashed line) are integrated over $0.8<h<1.2$ and 
$-0.1<k<0.1$. Data above the dashed line have been integrated over $-1<h<4$ and 
$|k|<1$ and intensities in this region have been multiplied by 4 to make the 
faint diffuse feature around $l=-1$ at about 5~meV more clearly visible. The ED 
calculations have been convolved with Gaussians of FWHM 0.066~meV, 0.16~meV and 
0.3~meV in B-D, F-H and J-L respectively. In B-D, the intensity shown is 
$S^{xx}(l,\omega)$, as the data in A have been corrected for the single ion 
magnetic form factor and the polarization factor under the assumption that 
$S^{xx}=S^{yy}$ and $S^{zz}=0$ for inelastic scattering. In F-H and J-L, all 
components of the dynamical structure factor are included and the integration 
range, polarization factor and magnetic form factor have all been accounted for 
in the calculation.}\label{F:dataEDcomparison}
\end{figure*}

The first term in $\mathcal{H}_1$ is the dominant Ising exchange. The second 
($\lambda_S$) is a symmetric XY exchange term, which allows single spin flips to 
hop, and the third term ($\lambda_{yz}$) is an off-diagonal staggered exchange 
term which allows domain walls to hop. These three terms (together with 
$\mathcal{H}_{\text{MF}}$, defined below, for zero field and transverse fields 
below the critical point) can qualitatively account for all of the features seen 
in the INS spectrum, not only in field parallel to the $a$-axis presented here, 
across the full range of field values considered, but also in purely transverse 
field ($\parallel b$) below \cite{Lovas2023} and above the critical field 
\cite{Fava2020}. In particular, the XY exchange is needed to account for the 
kinetic bound state seen in zero field (near $l=-1$ in 
Fig.~\ref{F:dataEDcomparison}A) \cite{Coldea2010}, and for the large $m_1$ 
bandwidth seen in field parallel to the $a$-direction 
(Fig.~\ref{F:dataEDcomparison}E), as this term allows single spin flips to hop. 
The staggered off-diagonal exchange term $\lambda_{yz}$ is needed to account for 
the fact that domain walls can hop in zero applied field (seen in the fact that 
the bands around $l=0$ in Fig.~\ref{F:dataEDcomparison}A are dispersive) 
\cite{Fava2020}. The key features of the nearest-neighbour exchange are 
graphically 
illustrated in Fig.~\ref{F:exchange}: the major axis of the ellipsoids at the 
middle of 
each bond corresponds to the Ising exchange, the diameter in the perpendicular 
plane 
to the Ising direction illustrate the XY exchange $\lambda_S$, and the staggered 
orientation of the ellipsoids is due to the staggered exchange $\lambda_{yz}$. 
Finally, the last term in (\ref{E:Hmin}) is the Zeeman interaction with an 
applied magnetic 
field, where we have assumed that the $g$-tensor is diagonal in the $xyz$ axes. 

In order to quantitatively account for the full wavevector dependence of the 
data, however, other, subleading terms are needed, included under 
$\mathcal{H}_2$: the first term, $\lambda_{\text{AF}}$, is a 
next-nearest-neighbour antiferromagnetic Ising term \cite{Coldea2010,Fava2020} 
needed to account for the energy of the kinetic bound state. The second term, 
$\lafxy$, is an XY part to the next-nearest-neighbour term which accounts for 
the flattening of the bottom of the $m_1$ dispersion in high field along $a$ 
[Fig.~\ref{F:dataEDcomparison}E]. The third term ($\lambda_{A}$) is an asymmetry 
between the XX and YY exchanges, which is needed to account for the position and 
bandwidth of the $m_2$ dispersion in large field along $a$; without this term, 
the $m_2$ mode is too low in energy. These last two terms are only of order 
2-3\% of the Ising term and the necessity for their inclusion in the 
parameterization can be seen by comparing  Fig.~\ref{F:dataEDcomparison}E with 
G. The discrepancies illustrated in the latter between the empirical and 
calculated dispersions motivates our further refinement of the Hamiltonian with 
the inclusion of the $\lambda_A$ and $\lambda_{\text{AF}}^{xy}$ terms 
which were fixed to zero in \cite{Fava2020}. 

For completeness, we note that, as mentioned in \cite{Fava2020}, two other 
nearest-neighbour 
exchange terms are symmetry allowed, 
$\lambda_{xz}J\sum_j\left(S_j^xS_{j+1}^z+S_j^zS_{j+1}^x\right)$ 
and $\lambda_{xy}J\sum_j(-1)^j\left(S_j^xS_{j+1}^y+S_j^yS_{j+1}^x\right)$. 
However, the definition of the axes used so far --- that $z$ is the direction of 
the equilibrium spin in zero field ---  places a constraint between 
these two terms, i.e., only one can vary independently. This is because each of 
these two terms on their own when added to (\ref{E:propHam}) leads to a rotation 
of the zero-field equilibrium spin direction away from the $z$-axis, but for a 
given $\lambda_{xz}$ one can choose a corresponding $\lambda_{xy}$ of 
appropriate magnitude and sign such that, when both terms are present, the 
zero-field equilibrium spin direction is still along $z$. However, we find that 
allowing finite $\lambda_{xz}$ and $\lambda_{xy}$ with the above constraint does 
not measurably improve the agreement with the present experimental data, so in 
the following we assume $\lambda_{xz}=\lambda_{xy}=0$. 

Finally, $\mathcal{H}_{\text{MF}}$ captures the effects of interchain couplings 
in a mean-field approximation, where in zero and low transverse field
\begin{equation}
\mathcal{H}_{\text{MF}} =-J \sum_j 2\lambda_{\text{MF}}\langle \mathbf{S}\rangle 
\cdot \mathbf{S}_j .\label{E:Hmf}
\end{equation}
In fields above 0.14~T applied along the $a$-direction, the relevant form is 
instead given in Sec.~\ref{S:otherfields}, and, at high field, this simplified 
form is no longer sufficient as excitations acquire a finite interchain 
dispersion at first order in the interchain couplings. Therefore, in this high 
field regime, we use the full explicit form for the relevant interchain 
exchanges proposed in (\ref{E:Hinterchain}).

\begin{table}
\centering
\begin{tabular}{cd}
\hline
\hline
$J$&2.48(2)\quad\text{meV}\\
$\lambda_S$&0.251(6)\\
$\lambda_A$ & -0.021(1)\\
$\lambda_{yz}$ & 0.226(3)\\
$\lambda_{\text{AF}}$ & 0.077(3)\\
$\lafxy$ & 0.031(1)\\
$g_x$ & 3.29(6)\\
$g_y$ & 3.32(2)\\
$g_z$ & 6.90(5)\\
$\lambda_{\text{MF}}$ & 0.0158(2)\\
\hline
\hline
\end{tabular}
\caption{Single-chain Hamiltonian parameters used in this work as defined in 
(\ref{E:Hmin})-(\ref{E:Hmf}).}
\label{T:hamparams}
\end{table}

The refined parameter values are shown in Table~\ref{T:hamparams}.  The 
Hamiltonian shown above quantitatively reproduces the spectrum seen in zero 
field, high field $\parallel~a$ and high field $\parallel~b$. Furthermore, it 
also reproduces data at low fields along $b$, as shown in \cite{Lovas2023} and 
at intermediate fields along $a$, and also accounts for previously published THz 
spectroscopy 
data in low transverse field, as we discuss later in Sec.~\ref{S:otherfields}.

\subsection{Fitting procedure}\label{S:singlechainfitting}

\begin{figure}
\includegraphics[width=0.3\textwidth]{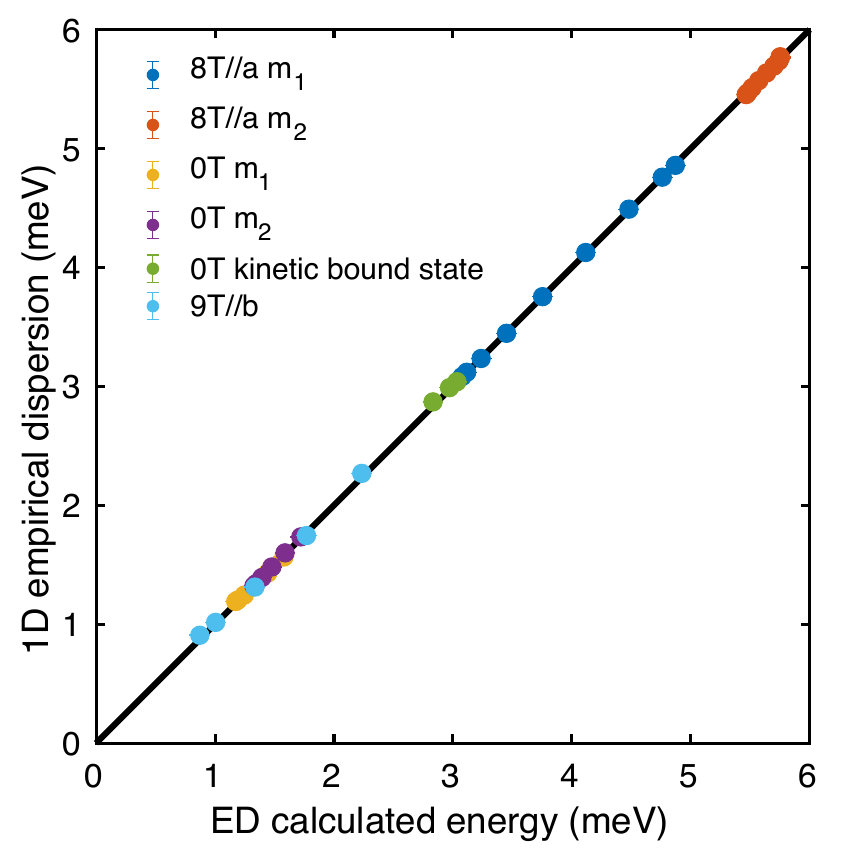}
\caption{Energies of dispersion points as calculated from ED on 16 sites 
compared to their values as calculated from the empirical dispersion relations. 
The 
one-to-one agreement obtained is excellent.}\label{F:EDobsvscalc}
\end{figure}

The fitting procedure used a global simultaneous fit to the dispersion relations 
corresponding to data taken in zero field, 8~T~$\parallel~a$ and 
9~T~$\parallel~b$, with the aim of arriving at a consistent description of all 
these different regimes within the same Hamiltonian. 

First, cuts were taken through the data as a function of energy transfer at 
constant wavevector transfer. Dispersion points were obtained by fitting 
Gaussian peak shapes to these cuts. Many dispersion points were extracted from 
each data set (over 500 for the 8~T~$\parallel~a$ data). Empirical dispersion 
relations were then fit to these dispersion points. For the high field data, 
these 
were 3D dispersion relations, as per (\ref{E:hopping}) and 
(\ref{E:hightransversedispersion}) for field along $a$ and $b$ respectively. In 
zero field, the interchain hopping effects are negligible, due to the multi-spin 
nature of the bound states as well as the antiferromagnetic order pattern 
between chains, which suppresses interchain hoppings for the kinetic bound state 
(see Appendix \ref{S:lowfieldhopping}) so a 1D form was used.
The single-chain Hamiltonian was then fit to these empirical dispersion 
relations 
with the interchain parameters set to zero. It was not possible to use the 
parameters derived from linear spin wave theory fits directly, since the 
predominantly 1D nature of the magnetic interactions together with the small 
effective spin $S=1/2$ leads to strong quantum fluctuations that renormalize the 
dispersions even at the high magnetic fields investigated. In addition, we 
wanted to capture the various bound states, i.e. the zero field confinement 
bound states, as well as the high field $\parallel a$ $m_2$ bound state; this is 
not possible in linear spin wave theory. Instead, exact diagonalization (ED) 
calculations on finite chains were used, with periodic boundary conditions at 
the ends of the chains. The fits used calculations on 12 sites as the best 
compromise between minimal finite size effects and the computation being quick 
enough to be carried out many times for fitting (for fitting to the kinetic 
bound state, 10 sites were used as this required more eigenstates to be found 
and is not strongly affected by finite size effects). Lanczos algorithms were 
used to diagonalize only the low energy subspace and speed up the calculation. 
There were 10 parameters in the fit but 13 pieces of information across these 
different data sets, so the fit was not underconstrained. In particular, the 
pieces of data used for the fit were the dispersions of the first two 
confinement bound states and the kinetic bound state from the zero field data 
\cite{Coldea2010}, the $m_1$ and $m_2$ dispersions from the 8~T $\parallel a$ 
data, and the magnon dispersion from the 9~T $\parallel b$ data. For each of 
these data sets, the squared difference $\chi^2$ between the ED at each momentum 
point and the empirical fitted dispersion relation was calculated, normalized by 
the 
uncertainty on the fitted dispersion relation as calculated from the covariance 
matrix of 
the fitted dispersion parameters, and summed; the fit minimized this total 
$\chi^2$. When calculating $\chi^2$ for the high transverse field data, only the 
portion of the dispersion for $l\leq0.5$ was used, as at higher values of $l$, 
quasiparticle breakdown occurs \cite{Robinson2014nw,Fava2020}, meaning that the 
dispersion relation ceases to be well defined. The optimization used a 
quasi-Newton 
algorithm and the initial parameters were varied in order to make sure that a 
global minimum was found.  The very good one-to-one agreement obtained in the 
simultaneous fits of the ED to the empirical dispersion relations is shown in 
Fig.~\ref{F:EDobsvscalc}. 

Uncertainties on the fitted parameters were estimated by varying the parameters 
of the empirical dispersion relations according to their covariance matrices.
There are significant correlations between some of the parameters, especially 
between $J$ and $\lambda_S$ and between $\lambda_S$ and $g_x$. However, while a 
number of slightly different parameter sets give similar agreement with the 
features to which the single-chain Hamiltonian was fit, the final parameter set 
presented in Table ~\ref{T:hamparams} gives the best agreement not only with the 
features to which the Hamiltonian was fit, but also to other features in the 
spectra.  These include the relative intensities of different features, the full 
bandwidth of the magnon in the high transverse field data, the energy and 
intensity 
of the faint diffuse feature around $l=-1$ at about 5~meV in the high transverse 
field data, the field dependence of the bandwidths of the $m_1$ and $m_2$ modes 
in field along $a$, and the spectra in low transverse field (see 
\cite{Lovas2023}).

A comparison between the data to which the Hamiltonian was fit and the spectrum 
calculated by exact diagonalization using this refined Hamiltonian is shown in 
the first two columns of Fig.~\ref{F:dataEDcomparison}. In these plots, the 
color indicates the measured/calculated scattering intensity, while the overlaid 
white dots/curves show the data points/dispersion relations that were being fit 
to. It 
can be seen that very good quantitative agreement is achieved. 
In these calculations, the plotted intensities are 
\begin{equation}\label{E:Sqw}
S(\mathbf{Q},\omega)=|f(Q)|^2\sum_{\alpha, 
\beta}\left(\delta_{\alpha,\beta}-\hat{Q}_\alpha\hat{Q}_\beta\right)S^{\alpha
\beta}
\end{equation}
where $\alpha, \beta$ both run over $x, y, z$, $f(Q)$ is the magnetic form 
factor, and $\hat{Q}_\alpha$ is the component along direction $\alpha$ of the 
unit vector parallel to the wavevector transfer $\mathbf{Q}$. The partial 
dynamical structure factors $S^{\alpha\beta}$ are
\begin{equation}\label{E:Sab}
S^{\alpha\beta}=g_{\alpha}g_{\beta}\sum_{\lambda_f}\langle 
\text{GS}|S^{\alpha}(\mathbf{Q})|\lambda_f\rangle\langle\lambda_f|S^{\beta}
(\mathbf{Q})|\text{GS}\rangle 
\delta(E_{\lambda_f}-\hbar\omega)
\end{equation}
where the sum is over all excited states $|\lambda_f\rangle$ with energy 
$E_{\lambda_f}$ relative to the ground state $|\text{GS}\rangle$ and where the 
Fourier transformed spin is 
$S^\alpha(\mathbf{Q})=\sum_{\mathbf{r}} S^\alpha_{\mathbf{r}} 
e^{i\mathbf{Q}\cdot\mathbf{r}}$, summing over all sites $\mathbf{r}$.

The right hand columns of Fig.~\ref{F:dataEDcomparison} contain comparisons to 
previous models that cannot quantitatively or qualitatively account for key 
features in the experimental data. These are discussed in 
Appendix~\ref{S:restrictedcomparisons}. This comparison provides evidence that 
all of the parameters are indeed needed in order to quantitatively capture all 
features in the data.

\subsection{Comparison of the Hamiltonian to experiment at other 
fields}\label{S:otherfields}

\begin{figure}
\includegraphics[width=0.5\textwidth]{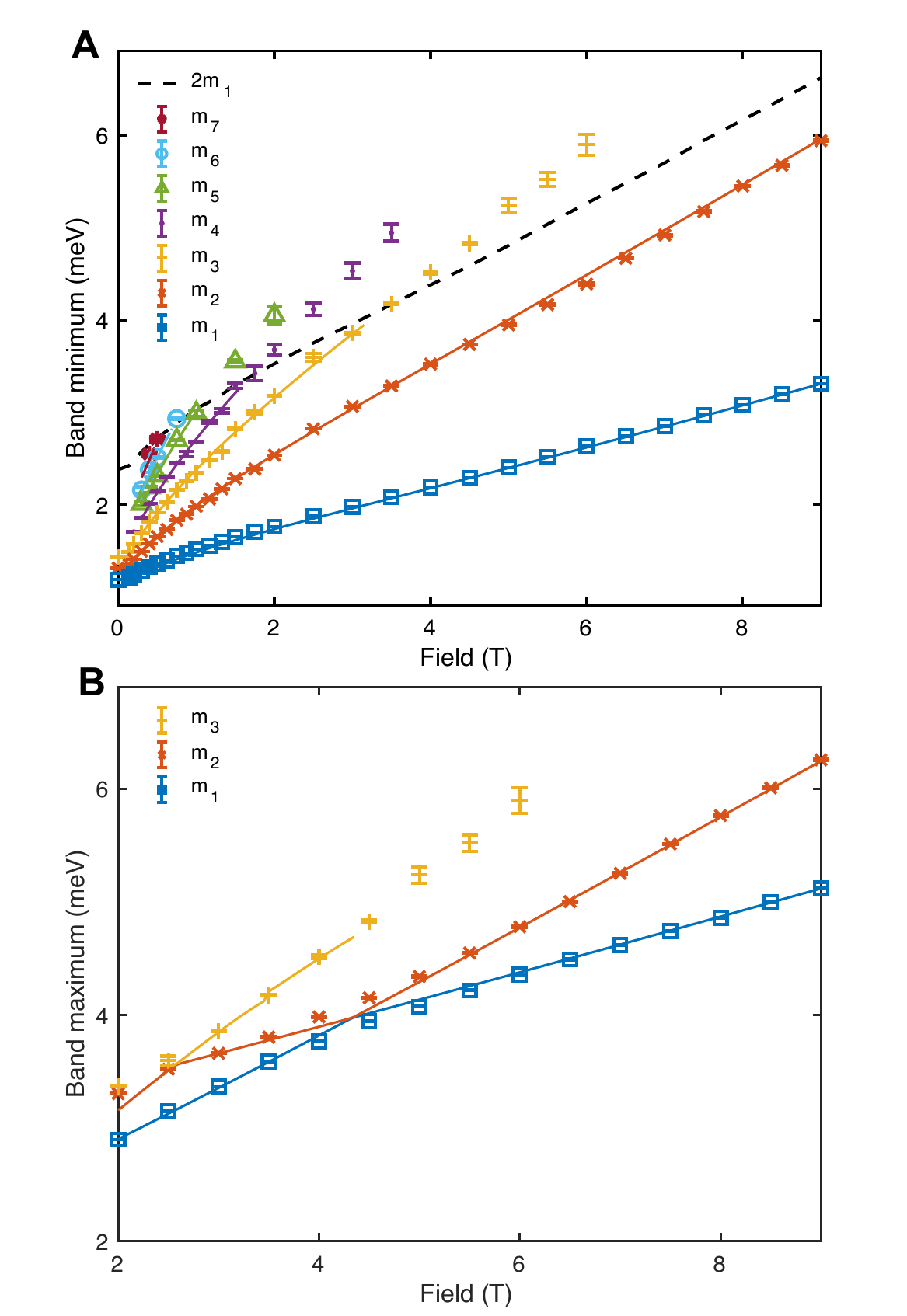}
\caption{Band minima (A) and maxima (B) showing experimentally extracted points 
(symbols) corrected for interchain dispersion. The results of ED calculations on 
16 sites (solid lines) are overlaid in the corresponding color, with effects of 
interchain couplings included as a mean-field correction. There is good 
agreement over the whole field range probed. The dependence of the band maxima 
on field was found to discriminate more strongly between different models than 
the field dependence of the band minima. Note that the higher bound states cease 
to be well defined in ED once they overlap with the continuum; for this reason 
solid lines stop when they intersect the dashed line indicating the lower 
boundary of the $2m_1$ continuum in A. In B, the discrepancy between ED and 
experiment where the $m_1$ and $m_2$ ED curves touch is due to the fact that 
experimental values were extracted by fitting to the hopping form 
(\ref{E:hopping}) and points were not extracted in regions where different modes 
overlapped, as they do in Fig.~\ref{F:strongconfinement}C and D. The interchange 
in the gradient of the maxima of  $m_1$ and $m_2$ modes with respect to field 
where they intersect is due to the exchange in character of the top of the mode: 
between approximately 2~T and 4~T, it is the top of the $m_2$ mode that has 
single-spin-flip character around $l=1$. 
}\label{F:allbandminimaED}
\end{figure}

\begin{figure}
\includegraphics[width=0.4\textwidth]{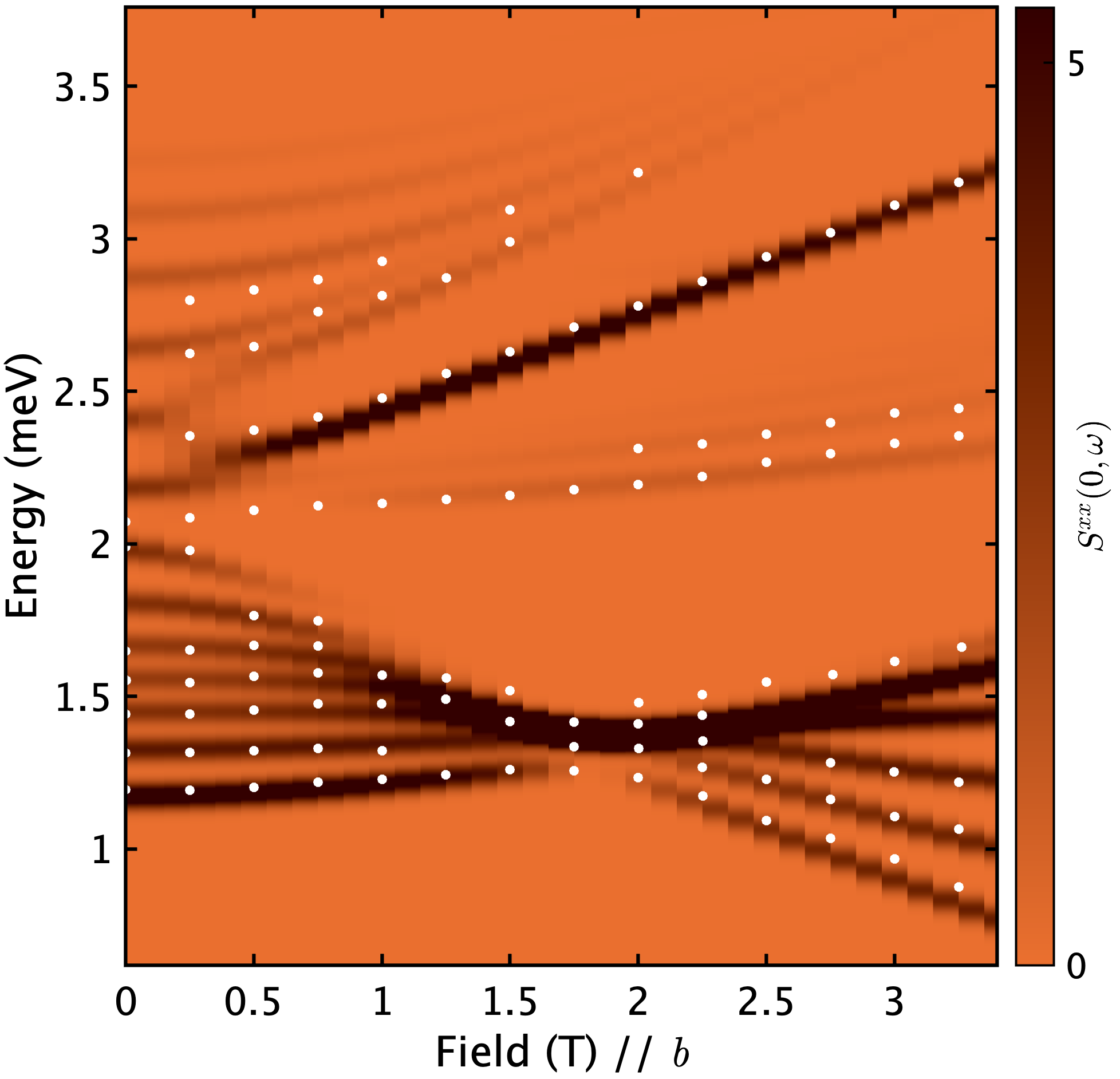}
\caption{
Dynamical structure 
factor at zero momentum transfer as a function of transverse field.
Color indicates $S^{xx}(0,\omega)$ as per (\ref{E:Sab}),  calculated using ED for a chain of
16 sites for the Hamiltonian in (\ref{E:propHam}) and convolved 
with a Gaussian of FWHM 0.067 meV. 
White dots are data points extracted from the estimated local intensity maxima 
in 
the energy-dependent THz spectroscopy data presented in \cite{Morris2021} at 
1.5~K (Fig. 2c of that work). 
}\label{F:ThzED}
\end{figure}

The refined Hamiltonian can also be compared to the results of INS data taken at 
fields other than those to which the fit was performed. The data agree 
significantly 
better with the calculation when using the fit with all terms included than when 
omitting any of the terms in $\mathcal{H}_2$.

Fig.~\ref{F:allbandminimaED} shows comparisons between ED calculations and the 
data at fields away from those used for the fits. Excellent quantitative 
agreement is found.
In the calculations shown here, the component of interchain interactions 
parallel to the magnetization direction was taken into account in a mean field 
picture.  
For the field-polarized phase in field above 0.14~T along the $a$-direction, the 
form of the interchain mean field in (\ref{E:Hmf}) does not hold because the 
ordering 
pattern of the chains changes, and instead the relevant form of the mean field 
term is
\begin{equation}\label{E:Hmf_a}
\mathcal{H}_{\text{MF}}= \sum_j 2\left(J_1+J_1^{\prime}\right)\langle 
\mathbf{S}\rangle \cdot \mathbf{S}_j - 4J_2\langle S^z \rangle S^z_j,
\end{equation}
where the interchain exchanges are defined in Appendix~\ref{S:interchaintrans}. 
This was the form used in Fig.~\ref{F:allbandminimaED}.

In addition, in Fig.~\ref{F:ThzED}, we compare recently reported THz 
spectroscopy data (white 
dots) with predictions based on the Hamiltonian proposed here. 
The comparison is shown for fields up to 3.5~T, which we consider to be in the 
region where ED is sufficiently reliable, as the gap is still large (i.e., 
finite size effects are small), and interchain effects are also small. All the 
trends in the THz spectroscopy data are reproduced. The good agreement in the 
intensities can 
be seen visually by comparing Fig.~\ref{F:ThzED} to Fig. 2c of \cite{Morris2021} 
(not shown).

\section{Conclusions}\label{S:conclusions}
In summary, we tuned the confinement potential between spinons in the 
ferromagnetic Ising chain material \ch{CoNb2O6} by applying an external magnetic 
field along the crystallographic $a$ direction, such that there was a large 
longitudinal (along Ising axis) field component. In the low field, weak 
confinement regime, we found a hierarchy of bound states, with their energy 
varying as field to the 2/3 power and the spacing between modes determined by 
the zeros of the Airy function, as expected in a picture of domain walls in a 
linear confining potential proportional to the strength of the longitudinal 
field. Upon increasing field, higher-order bound states increase more quickly in 
energy and progressively disappear from the spectrum such that in the limit of 
high field, in the strong confinement regime, we found only two bound states 
whose energies depend linearly on field. The higher energy of these two bound 
states is a dispersive two-spin-flip bound state which is stabilized by the 
proximity to the Ising limit. 
By performing a global fit to the full wavevector dependent spectrum observed in 
various fields, we also proposed a microscopic Hamiltonian including both 
in-chain and inter-chain interactions, down to 2\% of the dominant Ising 
exchange. This Hamiltonian quantitatively reproduces the INS data obtained 
across a wide range of field conditions including zero field, high 
near-longitudinal field and high transverse field, as well as intermediate 
fields to which the Hamiltonian was not fit, leading to a fully-consistent 
description of the spin dynamics using a single set of exchange parameters.  

\begin{acknowledgments}
L.W. acknowledges very useful discussions with Michele Fava. We thank Alexander 
Chernyshev and C\'{e}sar Gallegos for helpful comments and discussions. We 
acknowledge 
doctoral studentship funding from Lincoln College and the University of Oxford 
(L.W.), the Engineering and Physical Sciences Research Council (D.M.), and the 
University of Oxford Clarendon Scholarship Fund and NSERC of Canada (J.D.T). We 
acknowledge support from the Engineering and Physical Sciences Research Council 
grant numbers EP/H014934/1 (I.M.C. and R.C.) and GR/M47249/01 (D.P.). R.C. 
acknowledges support from the European Research Council under the European 
Union’s Horizon 2020 research and innovation programme Grant Agreement Number 
788814 (EQFT). R.C. also acknowledges support from the National Science 
Foundation 
under Grants No. NSF PHY-1748958 and PHY-2309135, and hospitality from KITP 
where 
part of this work was completed. The neutron scattering experiments at the ISIS 
Facility were supported by beamtime allocations from the Science and Technology 
Facilities Council. Figures~\ref{F:structure} and \ref{F:exchange} were made 
using the crystal visualization software \textsc{vesta} \cite{Momma2011}. Access 
to the data will be made available from Ref.~\cite{database}.
\end{acknowledgments}

\appendix

\section{Interchain dispersions in high field}\label{A:highfieldinterchain}

In this Appendix, we present the characterization of the experimentally observed 
interchain dispersion relations at high field and their quantitative 
parameterization. 

\subsection{Characterization of the interchain dispersions in high 
field}\label{S:charinterch}

\begin{figure*}
\includegraphics[width=\textwidth]{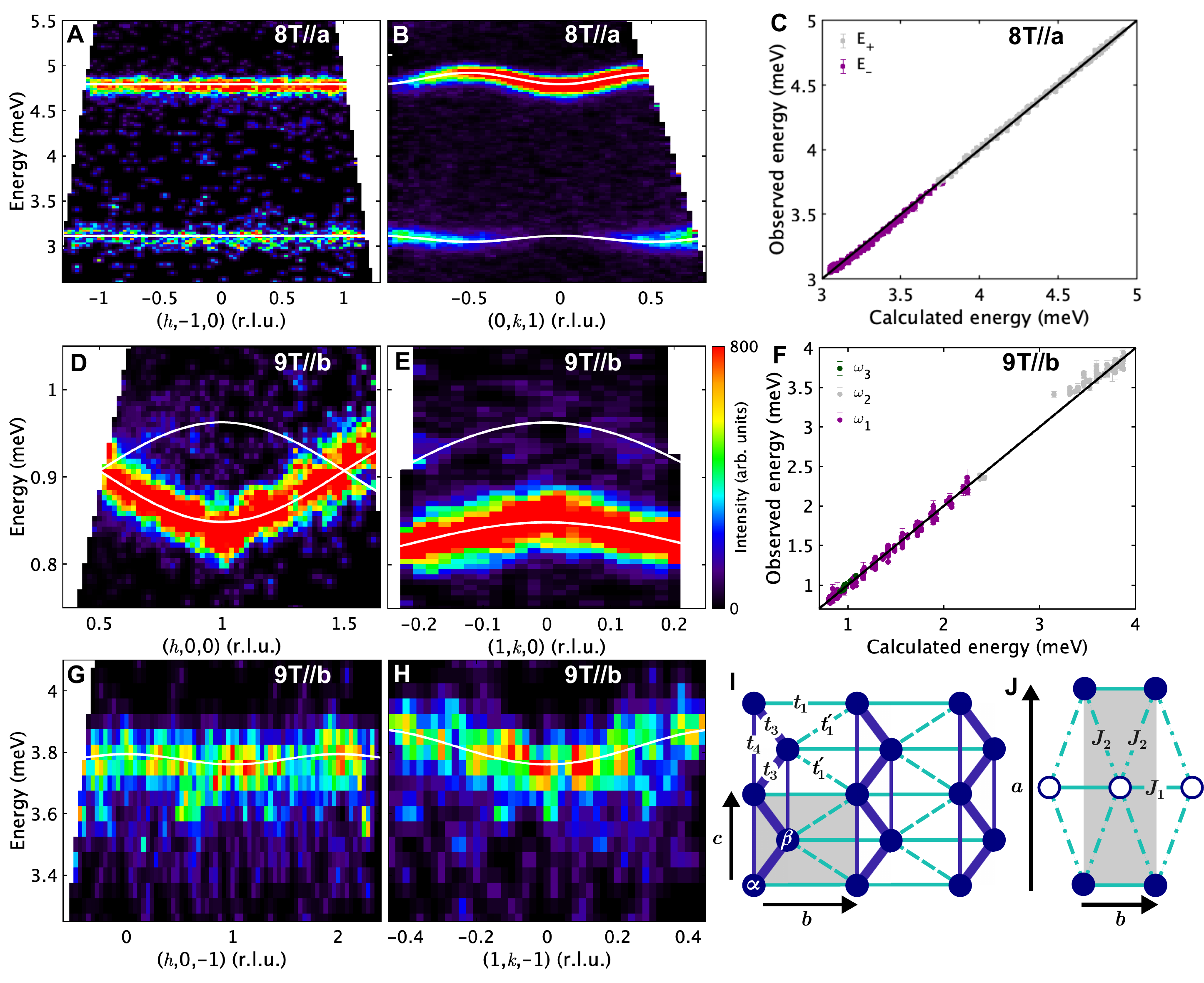}
\caption{
Interchain dispersions at 8~T~$\parallel a$ (A, B) and 9~T~$\parallel b$ (D, E, 
G, H). White curves represent best fit dispersion relations. Color is the raw 
neutron 
scattering intensity on an arbitrary scale. The incident energies $E_i$ used 
were 9.33~meV (A, B), 2.14~meV (D, E) and 10.17~meV (G, H), and the overall 
intensity scale is different for each experimental configuration.
C, F) Comparison of observed and calculated energies with best fit parameters 
for the 8~T~$\parallel a$ hopping dispersion (\ref{E:hopping}) and 
9~T~$\parallel b$ linear spin wave theory dispersion respectively, with 
$\omega_{1,2}(\mathbf{Q})=\omega_{\pm}(\mathbf{Q})$ and 
$\omega_3(\mathbf{Q})=\omega_-(\mathbf{Q}+\mathbf{a}^*)$, with $\omega_{\pm}$ as 
per (\ref{E:hightransversedispersion}).  I, J)  Exchange interaction paths in 
the $bc$ and $ab$ planes, respectively. Labels $t_{1,3,4}$ and $t_1^{\prime}$ 
are hopping parameters for the dispersion model in (\ref{E:hopping}), while 
$J_1$ and $J_2$ refer to the Hamiltonian (\ref{E:Hinterchain}). Dark/light 
colored lines indicate in-chain/inter-chain bonds, respectively. The gray shaded 
area indicates the structural unit cell shifted from the conventional $Pbcn$ 
unit cell such that the origin is at one \ch{Co^{2+}} site.
The site labels $\alpha$ and $\beta$ are used in the hopping calculations for 
the Hamiltonian in (\ref{E:Hhopping}) and in the linear spin wave theory 
calculations for the full interchain Hamiltonian in 
(\ref{E:Hinterchain}). 
Filled/open circles represent spins with the local Ising axis at an angle $\pm 
\gamma$ to the $c$ direction respectively. The experimental data were integrated 
in the transverse wavevector directions as follows: 
$-1.05\leq k\leq -0.95$, $-0.05\leq l \leq 0.05$ (A); $-1.5\leq h \leq 1.5$, 
$0.95\leq l \leq 1.05$ (B); $-0.1 \leq k \leq 0.1$, $-0.05\leq l \leq 0.05$ (D); 
$0.8\leq h \leq 1.2$, $-0.05\leq l \leq 0.05$ (E); $-0.1\leq k \leq 0.1$, 
$-1.05\leq l \leq -0.95$ (G); $0.8\leq h \leq 1.2$, $-1.05\leq l \leq -0.95$ 
(H).}\label{F:interchain}
\end{figure*}

INS data probing the dispersions in the interchain directions are plotted in 
Fig.~\ref{F:interchain}. Panel A shows that in large field along $a$, no 
systematic dispersion bandwidth 
is detected along $h$ within the experimental resolution 
used. Therefore, all data taken with field along $a$ presented in the rest of 
this work were integrated across the entire range of $h$ in the data, that is, 
over $-1.5<h<1.5$. Along $k$, however, there is visible dispersion and further, 
the bandwidth and even sign of the dispersion depends on the mode energy, i.e. 
on 
$l$. This is illustrated in Fig.~\ref{F:interchain}B, where the lower energy and 
higher energy modes have $k$ dispersions that are of opposite signs (white solid 
lines are fits as described in Appendix~\ref{S:interchainlong}). Note that the 
bottom mode is the shadow 
version of the primary $m_1$ mode wavevector-translated from $l=0$ whereas the 
top mode is the primary $m_1$ mode at $l=1$. This coupled $kl$ dispersion is 
accounted for by the exchange pathways shown schematically in 
Fig.~\ref{F:interchain}I and is discussed mathematically in the following 
subsection.

Panels D, E, G and H of Fig.~\ref{F:interchain}, meanwhile, show the dispersions 
in the 
directions perpendicular to the chains in a transverse field, 9~T~$\parallel~b$. 
It can be seen that in this regime, there is interchain dispersion along both 
$h$ and $k$, of similar magnitude to that seen along $k$ for $B~\parallel~a$. 
However, while the dispersion along $k$ remains of similar magnitude but 
switches sign when moving from $l=0$ (Fig.~\ref{F:interchain}E) to $l=-1$ 
(Fig.~\ref{F:interchain}H), i.e. from the bottom to the top of the dispersion, 
similarly to what is seen for $B~\parallel~a$, the dispersion along $h$ is 
strongly suppressed at $l=-1$ (Fig.~\ref{F:interchain}G) compared to at $l=0$ 
(Fig.~\ref{F:interchain}D).
The model proposed in \cite{Cabrera2014yj} captures some, but not all, of these 
interchain dispersion effects. In particular it
predicts that the interchain bandwidth is suppressed at higher energy (i.e. 
$l=-1$ compared to $l=0$). This is indeed seen along $h$, but not along $k$. The 
same model also predicts that there will be very little interchain dispersion 
for near-longitudinal field, which is again seen along $h$ but not along $k$; in 
fact, the dispersion bandwidth along $k$ at $l=1$ in near longitudinal field is 
approximately the same as in transverse field (the $l=0$ transverse field $k$ 
dispersion bandwidth is enhanced by the bonds in the $ab$-plane). 
We therefore expand and revise the interchain interaction model in 
\cite{Cabrera2014yj} with the proposal that the bonds with a component along the 
$a$-direction (the dash-dot light blue $J_2$ paths in Fig.~\ref{F:interchain}J) 
are approximately Ising, whereas those in the $bc$ plane [the solid and dashed 
light blue paths in Fig.~\ref{F:interchain}I, i.e., $J_1$ ($t_1$) and 
$J_1^{\prime}$ ($t_1^{\prime}$)] are approximately Heisenberg.

\subsection{Parameterization of the dispersions relations in high field}

Parameterization of the interchain dispersion relations was done by assuming 
that 
the dispersion of single spin flips in directions perpendicular to the chains 
could be quantitatively captured using a linear spin wave formalism, but with 
in-chain parameters which may be renormalized compared to those deduced from the 
exchange Hamiltonian. The linear spin wave formalism is 
expected to be asymptotically exact in the limit of high field (when the gap is 
much larger than the bandwidth). In this limit, the linear spin wave dispersion 
relation 
can be perturbatively expanded to become equivalent to a spin-flip quasiparticle 
hopping (or tight-binding) formalism. As the hopping formalism depends only on 
the exchange pathways, this same formalism can also be applied to quasiparticles 
composed of multiple spin-flips.
In the case of field along the $a$-direction, we seek to parameterize the 
dispersion of bound states of various different numbers of spin flips within a 
single formalism, and therefore use a hopping formalism. For high transverse 
field, we seek to parameterize only a single magnon mode, but in a case where 
the gap is fairly small compared to the bandwidth such that a perturbative 
expansion in terms of a hopping formalism does not hold; we therefore use a 
linear spin wave theory formalism.

\subsubsection{Hopping model parameterization of the 3D dispersions in high 
near-longitudinal field ($\parallel a$)}\label{S:interchainlong}

To account for the coupled $kl$ dispersion observed in high field $\parallel a$  
(see Fig.~\ref{F:interchain}B), we propose a hopping model, shown 
schematically in Fig.~\ref{F:interchain}I, where solid and dashed bonds indicate 
paths across which excitations could hop. 
This approach was used in order to be able to describe both the $m_1$ and the 
$m_2$ modes using the same formalism. The effects of the Ising parts of the 
interchain coupling are taken into account at a mean-field level (see 
Sec.~\ref{S:otherfields}) which affects both the $m_1$ and $m_2$ states. This 
more phenomenological formalism is also able to capture the full 3D 
dispersions of higher bound states in lower fields. In Fig.~\ref{F:interchain}I, 
the dark blue lines represent in-chain bonds (parameterized by $t_3$ and $t_4$), 
whose Hamiltonian is described in detail in (\ref{E:propHam}), while the light 
blue lines represent inter-chain bonds (parameterized by $t_1$ and 
$t_1^{\prime}$). The next-nearest-neighbour in-chain hopping ($t_4$) accounts 
for the flattening of the bottom of the dispersion [see Fig.~\ref{F:8Tdata}A] 
while the diagonal interchain bonds ($t_1^{\prime}$) explain why the dispersion 
bandwidth along $k$ depends on $l$.

We can write this hopping model for the paths illustrated in 
Fig.~\ref{F:interchain}I as a two sublattice model (a single zigzag chain with 
two sites per effective unit cell) with sublattices labeled $\alpha$ and 
$\beta$. The different Ising directions on the two chains in the structural unit 
cell can be neglected as we do not include interactions along the $a$ direction 
since no dispersion along $h$ is detected within the resolution 
of the experiment. This is consistent with the picture of the interchain 
couplings presented in the following subsection where the only coupling with a 
component along $a$, $J_2$, is Ising like. This means that the hopping induced 
by $J_2$ is suppressed by a factor of $\sin^2{\theta}$ where $\theta$ is the 
angle between the local Ising axis and the local magnetization direction, and is 
small even for large fields along $a$. Therefore, in high field along $a$, we 
can approximate the Hamiltonian as being decoupled into $bc$ planes with hopping 
interactions within planes and we write it as
\begin{align}
\mathcal{H}_{\text{hopping}}=\sum_{\mathbf{R}} 
&\omega_0\left(c^{\dagger}_{\mathbf{R},\alpha}c^{\phantom{\dagger}}_{\mathbf{R},
\alpha} 
+ c^{\dagger}_{\mathbf{R},\beta}c^{\phantom{\dagger}}_{\mathbf{R},\beta} \right) 
\nonumber \\
+&\biggl[ t_3 
\left(c^{\dagger}_{\mathbf{R},\alpha}c^{\phantom{\dagger}}_{\mathbf{R},\beta} + 
c^{\dagger}_{\mathbf{R},\beta}c^{\phantom{\dagger}}_{\mathbf{R}+\mathbf{c},
\alpha} 
 \right) \nonumber \\
+ &t_4 
\left(c^{\dagger}_{\mathbf{R},\alpha}c^{\phantom{\dagger}}_{\mathbf{R}+
\mathbf{c},\alpha}+c^{\dagger}_{\mathbf{R},\beta}c^{\phantom{\dagger}}
_{\mathbf{R}+\mathbf{c},\beta} 
\right) \nonumber \\
+ &t_1 
\left(c^{\dagger}_{\mathbf{R},\alpha}c^{\phantom{\dagger}}_{\mathbf{R}+
\mathbf{b},\alpha}+c^{\dagger}_{\mathbf{R},\beta}c^{\phantom{\dagger}}
_{\mathbf{R}+\mathbf{b},\beta} 
\right) \nonumber \\
+ &t_1^{\prime} 
\left(c^{\dagger}_{\mathbf{R},\beta}c^{\phantom{\dagger}}_{\mathbf{R}+
\mathbf{b},\alpha}+c^{\dagger}_{\mathbf{R},\beta}c^{\phantom{\dagger}}
_{\mathbf{R}+\mathbf{b}+\mathbf{c},\alpha} 
\right) \nonumber \\
+ &\text{Hermitian conjugate}\biggr]. \label{E:Hhopping}
\end{align}
where $c^{\dagger}_{\mathbf{R},\alpha}$ 
($c^{\phantom{\dagger}}_{\mathbf{R},\alpha}$) creates (annihilates) a 
quasiparticle on the $\alpha$ sublattice in the unit cell with origin at 
$\mathbf{R}$.
The sum runs over all single-zigzag-chain unit cells, with primitive lattice 
vectors $(\mathbf{a}-\mathbf{b})/2$, $\mathbf{b}$ and $\mathbf{c}$ (the 
projection of the primitive unit cell in the $ab$ plane is shown by the shaded 
area in Fig.~\ref{F:Isingaxes}B).
The dispersion relations in this two sublattice model are then obtained as
\begin{align}
E_{\pm}&=A\pm\left|B\right|,\nonumber\\
 A&=\omega_0+2t_4\cos 2\pi l+2t_1\cos 2\pi k, \nonumber\\
B&=2\cos \pi l\left(t_3e^{4\pi i k \zeta} + t_1^{\prime}e^{-2\pi 
i(1-2\zeta)k}\right).\label{E:hopping}
\end{align}
Here, $t_i$ are parameters of the model that ultimately originate from spin 
exchange interactions between sites connected by the bonds indicated in 
Fig.~\ref{F:interchain}I and $\zeta$ is the fractional distance in the 
$b$-direction of the \ch{Co^2+} ions from the centre of the zigzag. The $\pm$ 
signs are to be chosen for the primary/shadow modes respectively for $|l|<0.5$ 
and vice-versa for $0.5<l<1.5$. 
If the hopping quasiparticle is a spin flip, the dynamical structure factor in 
inelastic neutron scattering is expected to be proportional to
\begin{equation}\label{E:Ihopping}
I_{\pm}=1\pm\cos\left(\operatorname{arg}{B}\right).
\end{equation}

\begin{table}
\centering
\begin{tabular}{cd}
\hline
\hline
Parameter & \multicolumn{1}{r}{Value (meV)} \\
\hline
$2t_1$&-0.0147(4)\\
$2t_1^{\prime}$& 0.0477(6)\\
&\\
$J_1$ & -0.008(2)\\
$J_1^{\prime}$ & 0.040(2)\\
$J_2$ & 0.023(1)\\
&\\
$J\lambda_{\rm MF}$ & 0.0391(7)\\
\hline
\hline
\end{tabular}
\caption{
Interchain parameter values. $2t_{1}$, $2t_1^{\prime}$ are hopping parameters 
from the 8~T~$\parallel a$ fit to the $m_1$ dispersion. These should be compared 
to the values of $J_1$ and $J_1^{\prime}$ respectively. The fact that the 
respective values are comparable is evidence that these bonds are approximately 
Heisenberg. $J_1$, $J_1^{\prime}$ and $J_2$ are from the fit to 9~T~$\parallel 
b$ data. $J\lambda_{\rm MF}$ is the value of the zero-field interchain mean 
field derived from the fit to zero field data. We expect $J\lambda_{\rm 
MF}=J_1+J_1^{\prime}=0.032(1)$~meV, under the assumption that $J_1$ and 
$J_1^{\prime}$ are Heisenberg. This number is comparable to the fitted 
value.}\label{T:interchainparams}
\end{table}

The very good agreement between the observed and calculated $m_1$ dispersions
using the best fit parameter values is shown in Fig.~\ref{F:interchain}C and 
corresponding values for the interchain hopping parameters are listed in 
Table~\ref{T:interchainparams}. Similarly good agreement is found for the $m_2$ 
mode; in this case the interchain parameters $t_1$ and $t_1^{\prime}$ were set 
to zero because the $m_2$ state is a two-spin-flip state, and so any interchain 
dispersion would be second order in the interchain interaction strength, and 
thus expected to be too small to resolve experimentally.

The hopping model dispersion relation in (\ref{E:hopping}) was used to 
parameterize each 
mode at each field in the $B~\parallel~a$ data. One-dimensional band 
characteristics were obtained by setting $t_1$ and $t_1^{\prime}$ to zero and 
calculating the relevant characteristic based on the fitted values of the other 
parameters, that is $\omega_0$ (the energy offset due to Ising and Zeeman 
terms), $t_3$ and $t_4$. For the fits in Sec.~\ref{S:crossover}, these were the 
band average, $E_{\text{average}}=\omega_0$ and the band minimum, 
$E_{\text{min}}=\omega_0+2t_3+2t_4$ (note that $t_3$ is negative for all bands 
as the dominant nearest-neighbour interaction is ferromagnetic). It was found 
that, as the field decreases, the interchain dispersion decreases, which is 
consistent with the picture that upon lowering field, quasiparticles acquire a 
more pronounced multi-spin-flip character and the hopping of multi-spin-flip 
excitations is suppressed as it is of higher order in the interchain couplings. 
The kinetic bound state remains dispersive down to 0.14~T, consistent with this 
being a single-spin-flip mode, but has unresolvable dispersion in zero field, 
consistent with the zero field magnetic structure, in which the interchain 
hopping of the single spin flip is suppressed, as shown in 
Appendix~\ref{S:lowfieldhopping}. These trends were extracted from fits to the 
single orientation data, where the wavevector components $k$ and $l$ are 
coupled, and were confirmed by investigation of the 
Horace scan data set 
at 1.5~T.

We also note that the generic intensity from (\ref{E:Ihopping}) captures the 
general intensity dependence along the interchain directions as observed, e.g., 
in Fig.~\ref{F:interchain}B. Together with the good agreement with the 
dispersions, this justifies \textit{a posteriori} the use of a hopping model to 
parameterize the excitations in this regime.

\subsubsection{Linear spin-wave theory parameterization of the 3D dispersions at 
high transverse field ($\parallel b$)}\label{S:interchaintrans}

\begin{figure}
\includegraphics[width=0.45\textwidth]{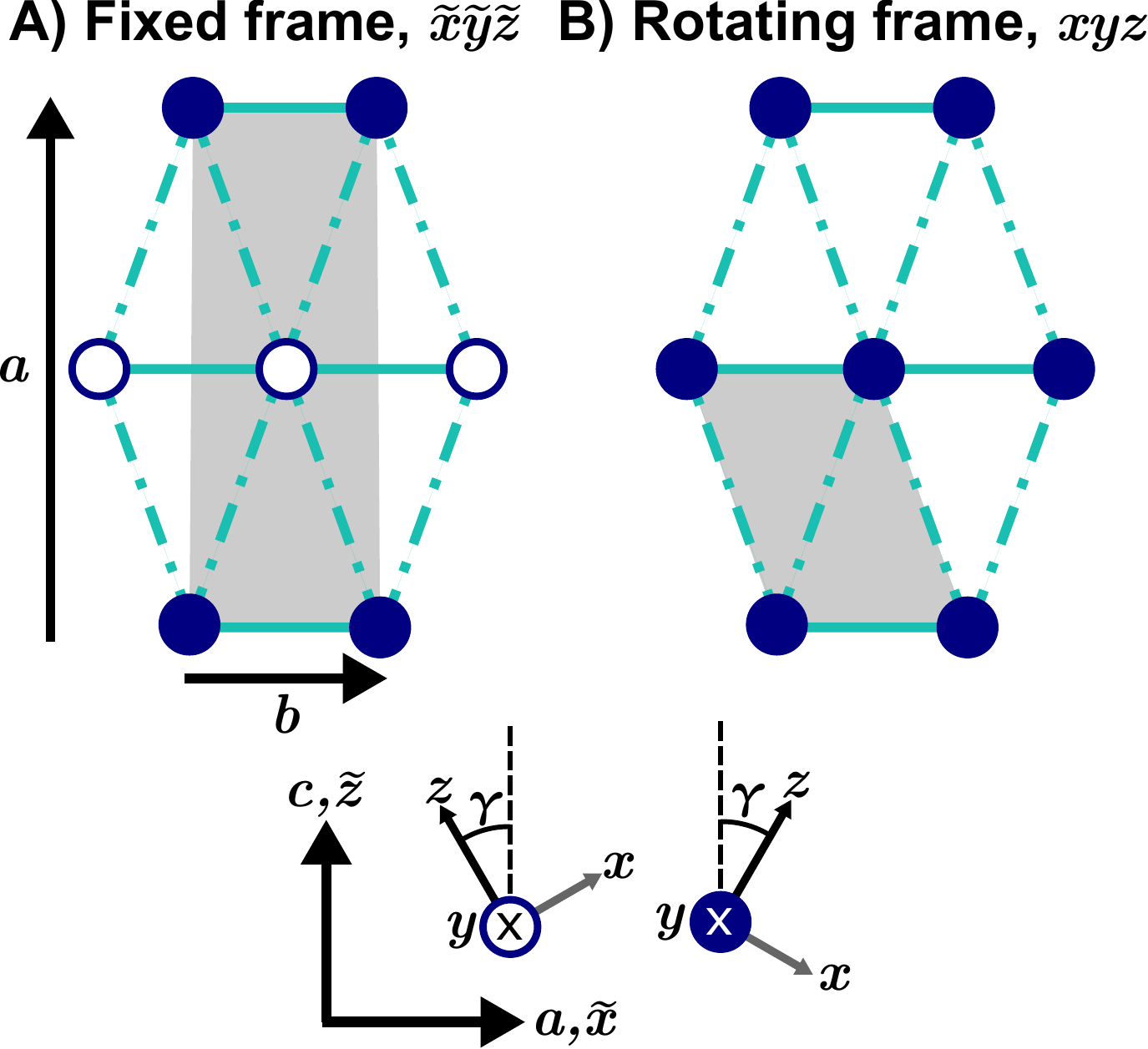}
\caption{A) $ab$ plane of the crystal structure where filled/open circles 
represent \ch{Co^{2+}} ions with the local Ising axis at an angle $\pm \gamma$ 
away from $\bm{c}$ towards $\bm{a}$. This alternation of the local Ising axis 
leads to two zigzag chains per orthorhombic unit cell (rectangular shaded area). 
These two chains are symmetry-related by both a $b$ glide (mirror in the 
$\tfrac{1}{4}\tilde{y}\tilde{z}$ plane followed by translation by $\bm{b}/2$) 
and an $n$-glide (mirror in the $\tilde{x}\tilde{y}\tfrac{1}{4}$ plane followed 
by translation by $(\bm{a}+\bm{b})/2$) of the $Pbcn$ space group of the crystal 
structure. Here $\tilde{x}\tilde{y}\tilde{z}$ define the fixed spin axes frame 
parallel to the orthorhombic $abc$ crystal axes. Bottom diagrams show definition 
of a mathematically convenient spin axes frame $xyz$ that rotates between the 
two chains such that $z$ is always along the local Ising axis. B) In this 
rotating spin axes frame the two chains become equivalent and the unit cell 
halves (shaded parallelogram).}\label{F:Isingaxes}
\end{figure}

We here discuss the formalism used to describe the interchain dispersion in high 
transverse field ($\parallel b$).

In the actual crystal structure of  \ch{CoNb2O6}, there are two zigzag chains 
per structural unit 
cell, shown in Fig.~\ref{F:Isingaxes}A (gray shading), one at the corner of the 
$ab$ cell (filled circle) with Ising axis tilted at an angle $+\gamma$ away from 
$c$ towards $a$, and one chain in the center (open circle) with Ising axis 
tilted at an angle $-\gamma$. In order to make progress analytically, we work in 
a reference frame where the Ising axes of the central chains are rotated to 
match those of the chains in the corners, such that the unit cell halves (as 
shown in Fig.~\ref{F:Isingaxes}B), reducing the problem from a four sublattice 
to a two sublattice problem. This is possible 
because the interchain couplings are assumed to have a simplified form, being 
Heisenberg or Ising-like. In this frame, the Hamiltonian is

\begin{align}
\mathcal{H}_{\text{total}}=\sum_{\text{chains}}&\mathcal{H}_{1}+\mathcal{H}_2 
\nonumber\\
+\sum_{\mathbf{R}}&J_1\left(\mathbf{S}_{\mathbf{R},\alpha}\cdot\mathbf{S}
_{\mathbf{R}+\mathbf{b},\alpha}+\mathbf{S}_{\mathbf{R},\beta}\cdot\mathbf{S}
_{\mathbf{R}+\mathbf{b},\beta}\right)\nonumber\\
+&J_1^{\prime}\left(\mathbf{S}_{\mathbf{R},\alpha}\cdot\mathbf{S}_{\mathbf{R}+
\mathbf{b},\beta}+\mathbf{S}_{\mathbf{R},\beta}\cdot\mathbf{S}_{\mathbf{R}-
\mathbf{b}+\mathbf{c},\alpha}\right)\nonumber\\
+&J_2\left[S^{z}_{\mathbf{R},\alpha}\left(S^{z}_{\mathbf{R}+(\mathbf{a}+
\mathbf{b})/2,\alpha}+S^{z}_{\mathbf{R}+(\mathbf{a}-\mathbf{b})/2,\alpha}\right) 
\right.\nonumber\\
&+\left.S^{z}_{\mathbf{R},\beta}\left(S^{z}_{\mathbf{R}+(\mathbf{a}+\mathbf{b})/
2,\beta}+S^{z}_{\mathbf{R}+(\mathbf{a}-\mathbf{b})/2,\beta}\right)\right]. 
\label{E:Hinterchain}
\end{align}
where $\mathcal{H}_{1,2}$ contain in-chain interactions defined in the 
(\ref{E:Hmin},\ref{E:H2}).
Here, $\mathbf{R}$ runs over all single zigzag chain unit cells, as in 
(\ref{E:Hhopping}).

We now solve this Hamiltonian in linear spin wave theory.
Assuming spins polarized along the $b$-axis, after a Holstein-Primakoff 
transformation
and a Fourier transform, the Hamiltonian in (\ref{E:Hinterchain}) takes the form 
(up to quadratic order in magnon operators and omitting the constant term) 
\begin{equation}\label{E:H_LSWT}
\mathcal{H}=\frac{1}{2}\sum_{\mathbf{Q}}\mathbf{\Upsilon}_{\mathbf{Q}}^{\dagger}
\mathcal{D}(\mathbf{Q})\mathbf{\Upsilon}_{\mathbf{Q}}
\end{equation}
where
$\Upsilon_{\mathbf{Q}}^{\dagger}=(a_{\mathbf{Q}}^{\dagger},b_{\mathbf{Q}}
^{\dagger},a_{-\mathbf{Q}},b_{-\mathbf{Q}})$ 
and $a_{\mathbf{Q}}^{\dagger}$ and $b_{\mathbf{Q}}^{\dagger}$ are the magnon 
creation operators for the two single-chain sublattices. The matrix 
$\mathcal{D}(\mathbf{Q})$ has the form
\begin{equation}\label{E:D_LSWT}
\mathcal{D}(\mathbf{Q})=\begin{pmatrix} 
A&B&C&D^*\\
B^*&A&D&C\\
C^*&D^*&A&B\\
D&C^*&B^*&A
\end{pmatrix},
\end{equation}
where $A$, $B$, $C$ and $D$ are functions of $\mathbf{Q}$ and are given by
\begin{align}
A=&J\left(\lambda_S-\lambda_A-\lafxy\right)-J_1-J_1^{\prime}+g_y\mu_BB
\nonumber\\
+&J\frac{\lambda_{\text{AF}}+\lafxy}{2}\cos{2\pi l}+J_1\cos{2\pi k}+J_2\cos{\pi 
h}\cos{\pi k}\nonumber\\
B=&\left[-J\frac{(\lambda_S+\lambda_A+1)}{2}+J_1^{\prime}e^{2\pi i 
k}\right]e^{-4\pi i\zeta k}\cos{\pi k}\nonumber\\
C=&J\frac{\lambda_{\text{AF}}-\lafxy}{2}e^{-2\pi il}+J_2e^{-\pi ih}\cos{\pi 
k}\nonumber\\
D=&J\frac{\lambda_S+\lambda_A-1}{2}e^{4\pi i\zeta k} \cos{\pi l}.
\end{align}
Note that the off-diagonal staggered exchange, $\lambda_{yz}$, does not appear 
in these expressions, since this term vanishes in the linear spin wave theory 
approximation when the ground state is fully aligned along the $y$
direction. However, this term leads to higher order interactions which cause 
quasiparticle breakdown in the region where this is kinematically allowed 
\cite{Robinson2014nw,Fava2020}. This effect is not captured in the linear spin 
wave treatment.  

The spin wave Hamiltonian in (\ref{E:H_LSWT}, \ref{E:D_LSWT}) has the same 
functional form as another two sublattice system discussed in detail in 
\cite{Elliot2021} so the derivation of the dispersion relations and dynamical 
correlation 
functions is identical and we only reproduce key steps here.

The dispersion relations are obtained by diagonalizing the matrix 
$\mathcal{G}\mathcal{D}(\mathbf{Q})$ where $\mathcal{G}=\diag(1,1,-1,-1)$ and 
are given by
\begin{widetext}
\begin{equation}\label{E:hightransversedispersion}
\omega_{\pm}^2=A^2+|B|^2-|C|^2-|D|^2\pm\sqrt{(2AB-CD^*-D^*C^*)(2AB^*-CD-C^*D)+
(BD-B^*D*)^2}.
\end{equation}
\end{widetext}

To calculate the neutron scattering intensities, we need the right eigenvectors 
of $\mathcal{GD}$. The components of these are
\begin{widetext}
\begin{align*}
W(\omega)=&-(A+\omega)(A^2+|B|^2-|C|^2-|D|^2-\omega^2)+2A|B|^2-BDC^*-B^*D^*C\\
X(\omega)=&(A^2C^*+|B|^2C^*-|C|^2C^*+|D|^2C-C^*\omega^2)-A(BD-B^*D^*)+\omega(BD-
B^*D^*)\\
Y(\omega)=&B^*\left[(A+\omega)^2-|B|^2+|C|^2\right]-(ACD+ADC^*+CD\omega+C^*D
\omega)+BD^2\\
Z(\omega)=&D(A^2+C^{*2}-|D|^2-\omega^2)+B^{*2}D^*-2AB^*C^*
\end{align*}
\end{widetext}
up to a normalization
\begin{equation*}
\mathcal{N}(\omega)=\sqrt{|-|W|^2+|X|^2-|Y|^2+|Z|^2|}.
\end{equation*}

From this we obtain, in the rotating frame,
\begin{align}
S^{xx}(\mathbf{Q},\omega)=&|W-X+Y-Z|^2 
\frac{[\delta^+(\mathbf{Q})+\delta^-(\mathbf{Q})]}{4\mathcal{N}}\nonumber\\
S^{zz}(\mathbf{Q},\omega)=&|W+X+Y+Z|^2 
\frac{[\delta^+(\mathbf{Q})+\delta^-(\mathbf{Q})]}{4\mathcal{N}}\nonumber\\
S^{yy}(\mathbf{Q},\omega)=&0. \label{E:LSWTintensities}
\end{align}
where $\delta^{\pm}(\mathbf{Q})=\delta(\omega-\omega_{\pm}(\mathbf{Q}))$.

We now transform to global coordinates $\tilde{x}\tilde{y}\tilde{z}$, 
defined to be parallel to the $abc$ crystallographic axes respectively, with the 
two local $z$ axes defined as 
$\hat{\mathbf{z}}_{\pm}=\tilde{\mathbf{z}}\cos{\gamma}\pm\tilde{\mathbf{x}}
\sin{\gamma}$ 
with the sign alternating between the layers along the $a$-direction as shown in 
Fig. \ref{F:Isingaxes}A.

 Thus we have
\begin{align*}
S^x(\mathbf{Q})=&\sum_{\mathbf{r}}(S^{\tilde{x}}\cos{\gamma}-e^{2\pi 
ir_{\tilde{x}}/a}S^{\tilde{z}}\sin{\gamma})e^{i\mathbf{Q}\cdot\mathbf{r}} \\
S^z(\mathbf{Q})=&\sum_{\mathbf{r}}(S^{\tilde{z}}\cos{\gamma}+e^{2\pi 
ir_{\tilde{x}}/a}S^{\tilde{x}}\sin{\gamma})e^{i\mathbf{Q}\cdot\mathbf{r}}\\
S^y(\mathbf{Q})=&\sum_{\mathbf{r}}S^{\tilde{y}}e^{i\mathbf{Q}\cdot\mathbf{r}}
\end{align*}
where $\mathbf{r}$ runs over all magnetic sites and where $r_{\tilde{x}}$ is the 
component of $\mathbf{r}$ along $\tilde{x}$. 
Transforming the other way,
\begin{align*}
S^{\tilde{x}}(\mathbf{Q})=&\sum_{\mathbf{r}}(S^{x}\cos{\gamma}+e^{2\pi 
ir_{\tilde{x}}/a}S^{{z}}\sin{\gamma})e^{i\mathbf{Q}\cdot\mathbf{r}} \\
S^{\tilde{z}}(\mathbf{Q})=&\sum_{\mathbf{r}}(S^{{z}}\cos{\gamma}-e^{2\pi 
ir_{\tilde{x}}/a}S^{{x}}\sin{\gamma})e^{i\mathbf{Q}\cdot\mathbf{r}}\\
S^{\tilde{y}}(\mathbf{Q})=&\sum_{\mathbf{r}}S^{y}e^{i\mathbf{Q}\cdot\mathbf{r}}
\end{align*}
such that
\begin{align*}
S^{\tilde{x}}(\mathbf{Q})=&S^x(\mathbf{Q})\cos{\gamma}+S^z(\mathbf{Q}+\mathbf{a}
^*)\sin{\gamma}\\
S^{\tilde{z}}(\mathbf{Q})=&S^z(\mathbf{Q})\cos{\gamma}-S^x(\mathbf{Q}+\mathbf{a}
^*)\sin{\gamma}.
\end{align*}
Thus the Fourier transformed magnetization components are (in units of $\mu_B$)
\begin{align*}
M^{\tilde{x}}(\mathbf{Q})=&g_xS^{x}(\mathbf{Q})\cos{\gamma}+g_zS^{z}(\mathbf{Q}+
\mathbf{a}^*)\sin{\gamma}\\
M^{\tilde{z}}(\mathbf{Q})=&g_zS^{z}(\mathbf{Q})\cos{\gamma}-g_xS^{x}(\mathbf{Q}+
\mathbf{a}^*)\sin{\gamma}.
\end{align*}
From this, we get the dynamical structure factor including the neutron 
polarization factors as
\begin{align}
&S(\mathbf{Q},\omega)=\nonumber\\
&\left(1-\frac{Q_{\tilde{x}}^2}{\mathbf{Q}^2}\right)\left(g_x^2S^{xx}
(\mathbf{Q},\omega)\cos^2{\gamma}+g_z^2S^{zz}(\mathbf{Q}+\mathbf{a}^*,\omega)
\sin^2{\gamma}\right)\nonumber\\
&+\left(1-\frac{Q_{\tilde{z}}^2}{\mathbf{Q}^2}\right)\left(g_z^2S^{zz}
(\mathbf{Q},\omega)\cos^2{\gamma}+g_x^2S^{xx}(\mathbf{Q}+\mathbf{a}^*,\omega)
\sin^2{\gamma}\right),\label{E:Skw}
\end{align}
where $Q_{\tilde{x},\tilde{z}}$ are the components of $\mathbf{Q}$ projected 
along $\tilde{x}$ and $\tilde{z}$ respectively.
In this expression, there are no mixed polarization terms as $S^{xz}=-S^{zx}$, 
and 
there are no cross-terms involving Fourier transformed spins at different 
wavevectors since these are zero by conservation of momentum.

To obtain intensities to compare with experiment, the delta functions in 
(\ref{E:LSWTintensities}) are replaced by Gaussians in energy to reflect the 
finite instrumental energy resolution and the intensity is multiplied by the 
squared spherical magnetic form factor for \ch{Co^{2+}}. The analytic 
calculations 
presented here were cross-checked against numerical calculations performed using 
SpinW \cite{Toth2015nu}. 

The fact that the intensity contains terms with shifted wavevector 
$\mathbf{Q}+\mathbf{a}^*$ is a mathematical expression of the real space unit 
cell doubling (due to alternation of Ising axes) leading to Brillouin zone 
folding and a consequent shadow mode. This shadow mode is responsible for the 
additional weak scattering intensity at slightly higher energy than the main 
mode in Figs.~\ref{F:interchain}D and E. We see from (\ref{E:Skw}) that the 
intensity of the shadow mode relative to the primary mode is proportional to 
$\tan^2{\gamma}$. Note that this shadow mode is distinct from the shadow mode 
due to the zigzag of the chains, such that the full model in the fixed frame has 
a total of four dispersion relations, $\omega_{\pm}(\mathbf{Q})$ and 
$\omega_{\pm}(\mathbf{Q}+\mathbf{a}^*)$. 

In order to extract a parameterization of the 3D dispersion in high transverse 
field, fits were done to the linear spin wave dispersion relations 
(\ref{E:hightransversedispersion}).  The fitted values for the interchain 
parameters are shown in Table \ref{T:interchainparams}. We note that these 
parameters yield minima in the magnon dispersion relations at $(1,\pm q,0)$ 
with $q \sim 0.39$ which is compatible with the minima observed at 7~T 
$\parallel b$ in 
Fig.~5 of \cite{Cabrera2014yj}, and can thus be related to the propagation 
vector $q = 0.37$ of the spontaneous incommensurate spin density wave order that 
sets in 
at the magnetic ordering temperature 2.95~K in zero field \cite{Heid1995bt}.
To draw a direct comparison with interchain exchange parameters proposed 
in Ref. \cite{Cabrera2014yj}, the relevant value to compare to $J_1$ in 
\cite{Cabrera2014yj} 
is $J_1 + J_1^{\prime}$ in the present work.

\section{Suppression of the interchain hopping of the kinetic bound state in 
zero field}\label{S:lowfieldhopping}

A surprising result found experimentally is that the kinetic bound state, the 
sharp mode near $l=-1$ at the top of the spectrum in zero field 
(Fig.~\ref{F:dataEDcomparison}A), has no measurable dispersion in either 
interchain direction, even though it is a single-spin-flip state, and 
generically one expects interchain dispersion at first order in the interchain 
couplings. We explain in this Appendix that this is due to the combination of 
the particular antiferromagnetic order pattern between chains and the form of 
the interchain couplings by introducing a simplified toy model.

As the interchain $J_2$ bonds in Fig.~\ref{F:interchain}J are Ising like, and in 
zero field the spins are aligned along the Ising axis, the $J_2$ terms create no 
hopping along $h$. The $J_2$ bonds also do not contribute to a mean field, as 
the triangular lattice and the antiferromagnetic pattern mean that the effects 
of the two bonds cancel out. We therefore only need to consider the interchain 
bonds in the $bc$ plane.
In the simplest approximation, we consider straight ($\zeta=0$) ferromagnetic 
Ising chains along the $c$-direction, with an antiferromagnetic Heisenberg 
interaction between chains in the $b$ direction,
i.e.,
\begin{equation*}
\mathcal{H}_{\text{simple}}=\sum_{\mathbf{r}}-JS^z_{\mathbf{r}}S^z_{\mathbf{r}+
\mathbf{c}/2}+J_1\mathbf{S}_{\mathbf{r}}\cdot\mathbf{S}_{\mathbf{r}+\mathbf{b}}
\end{equation*}
where the sum is over all magnetic sites.
In the absence of any applied field, the chains order in an antiferromagnetic 
pattern along the $b$-direction (both in this simplified model and in the real 
material). Therefore, to calculate the linear spin wave spectrum, the 
quantization axis must be rotated by $180^{\circ}$ about the $y$-direction on 
alternating chains along the $b$-direction such that the ordered spin is along 
the $+z$ direction on all chains. In this rotating frame, the quadratic spin 
wave Hamiltonian is  
\begin{equation*}
\frac{1}{2}\sum_{\mathbf{Q}}\mathbf{\Upsilon}_{\mathbf{Q}}^{\dagger}\mathcal{D}
(\mathbf{Q})\mathbf{\Upsilon}_{\mathbf{Q}}
\end{equation*}
where 
$\Upsilon_{\mathbf{Q}}^{\dagger}=(a_{\mathbf{Q}}^{\dagger},a_{-\mathbf{Q}})$ and 
\begin{equation*}
\mathcal{D}(\mathbf{Q})=\begin{pmatrix}
J+J_1 & J_1\cos{2\pi k}\\
J_1\cos{2\pi k} & J+J_1
\end{pmatrix}.
\end{equation*}
This gives a dispersion relation
\begin{equation*}
\omega(\mathbf{Q})=\sqrt{(J+J_1)^2-J_1^2\cos^2{2\pi k}}.
\end{equation*}
Now, $J_1\ll J$ (experimentally, $J_1/J \lesssim 2\%$), so this can be Taylor 
expanded, such that the bandwidth along $k$ is proportional to $J_1^2/J$, i.e. 
appears at second order in $J_1$. This is not experimentally resolvable, which 
explains why no interchain dispersion is observed for the kinetic bound state in 
zero field. In contrast, in field $\parallel a$ above 0.14 T, the spin 
components along $a$ are all parallel, with the consequence that the interchain 
dispersion along $k$ appears at first order in $J_1$. This is experimentally 
resolvable, and indeed is clearly detected experimentally, with the kinetic 
bound state at 1.5~T having a bandwidth along $k$ of 0.114(2)~meV (not shown).

\section{Comparison to other Hamiltonian parameter 
sets}\label{S:restrictedcomparisons}

\begin{table}
\centering
\begin{tabular}{cc}
\hline
\hline
Parameters fixed & $\chi^2$\\
\hline
None & 2008.17\\
$\lambda_A=0$& 2718.21\\
$\lafxy=0$ & 5910.82\\
$\lambda_A=\lafxy=0$ & 7009.55\\
\hline
\hline
\end{tabular}
\caption{Values of $\chi^2$ corresponding to fixing some of the Hamiltonian 
(\ref{E:propHam}) parameters to zero}\label{T:chi2}
\end{table}

In this Appendix, we provide evidence that all the terms in the Hamiltonian 
(\ref{E:propHam}) are indeed needed in order to fit all features of the 
dispersive modes in the full data set.

The values of $\chi^2$ corresponding to fits in which some of the parameters in 
$\mathcal{H}_2$ are fixed to zero are shown in Table. \ref{T:chi2}. We only 
consider setting to zero those parameters which were not included in the 
parameterization of Ref. \cite{Fava2020}, as the parameters fit in that work 
were shown to all be necessary to parameterize just the zero field data. It is 
seen that it is not possible to fit all features in the data well without using 
all parameters. These values have not been divided by the number of data points 
because the fits were not done directly to the data but instead to the empirical 
1D dispersion relations corrected for interchain dispersion effects. However, 
the fit was 
indirectly performed to many hundreds of data points. 

In addition, Fig.~\ref{F:dataEDcomparison} shows comparisons between the data 
and ED calculations using different Hamiltonian models. The third column (panels 
C, G, K) presents calculations using the Hamiltonian model proposed in 
\cite{Fava2020}, with $g_x$ and $g_z$ taken from Table~\ref{T:hamparams}. The 
calculation for the 8~T~$\parallel a$ data (panel G) does not fully capture the 
flattening of the bottom of the $m_1$ dispersion and the calculated $m_2$ 
dispersion is shifted downwards from where it is found empirically. It is found 
that in order to capture the former effect, it is necessary to include either 
$\lafxy$ or $\lambda_A$ in the fits, and that in order to capture the position 
and shape of the $m_2$ dispersion, both of these must be non-zero.

The right-most column of Fig.~\ref{F:dataEDcomparison} presents calculations 
using the Hamiltonian model used in \cite{Morris2021}, which contains a subset 
of the nearest-neighbour exchange terms in (\ref{E:Hmin}) and (\ref{E:H2}) with 
certain constraints between the parameter values. While that model can capture 
the energy levels at the zone center ($l=0$) in zero and transverse field, i.e., 
the regime probed in \cite{Morris2021}, we find significant qualitative and 
quantitative discrepancies between calculations using that model and the full 
spectrum observed via INS. The model does not capture the wavevector dependence 
of the spectrum either in zero field, where the kinetic bound state near $l=-1$ 
is not captured at all [compare Fig.~\ref{F:dataEDcomparison}D (calculation) 
with A (data)], or in field applied along either $a$ or $b$, where the predicted 
magnon bandwidths are much smaller than observed experimentally (compare H and L 
with E and I, respectively). Moreover, in high field along $a$ (panel H), the 
model does not capture the spectrum even at $l=0$, as the small magnon 
bandwidth, which is underestimated by almost a factor of 2, leads to too large a 
predicted magnon gap. 
We ascribe these differences primarily to the fact that in the model used in 
\cite{Morris2021} there is no $S^x_jS^x_{j+1}$ exchange term and only a very 
small $S^y_jS^y_{j+1}$ term. In contrast, in the present work, we find that the 
$S^x_jS^x_{j+1}$ and $S^y_jS^y_{j+1}$ exchange terms are of very similar size, 
and of magnitude comparable to that of the staggered off-diagonal exchange, 
those being the main subleading terms after the dominant Ising exchange, as 
already noted in \cite{Fava2020}. The fully refined Hamiltonian model we propose 
in (\ref{E:propHam}) accounts quantitatively not only for the full energy and 
wavevector-dependence of the INS spectrum at all probed fields aligned along two 
orthogonal directions, but also for the THz spectroscopy data in 
\cite{Morris2021} without any adjustable parameters, as discussed in 
Sec.~\ref{S:otherfields}. Therefore, we propose that it is an accurate model of 
the actual spin Hamiltonian in CoNb$_2$O$_6$.

\end{document}